\documentclass[twocolumn,preprintnumbers,amsmath,amssymb]{revtex4-2}
\usepackage{amssymb}
\usepackage{amsmath} 
\usepackage{graphicx}
\usepackage{booktabs}
\usepackage{caption}
\usepackage{subcaption}
\usepackage{multirow}
\usepackage{verbatim}

\begin{document}
\title{On performance of thin-film meso-structured perovskite solar cell
through experimental analysis and device simulation}
\author{F. Bonn\'\i n-Ripoll$^1$}
\author{Ya. B. Martynov$^2$}
\author{R. G. Nazmitdinov $^{3,4}$}
\author{K. Tabah$^5$}
\author{C. Pereyra$^5$}
\author{M. Lira-Cantú$^5$}
\author{G. Cardona$^6$}
\author{R. Pujol-Nadal$^1$}

\affiliation{$^1$Departament d'Enginyeria Industrial i Construcci\'o,
Universitat de les Illes Balears, E-07122 Palma, Spain.}
\affiliation{$^2$State Scientific-Production Enterprise ``Istok"
Fryazino, Russia.}
\affiliation{$^3$Bogoliubov Laboratory of Theoretical Physics, Joint Institute for Nuclear
Research, 141980 Dubna, Russia.}
\affiliation{$^4$Dubna State University, 141982 Dubna, Moscow region, Russia.}
\affiliation{$^5$Catalan Institute of Nanoscience and Nanotechnology (ICN2), CSIC and the Barcelona
Institute of Science and Technology (BIST), Building ICN2, Campus UAB, E-08193 Bellaterra, Barcelona, Spain.}
\affiliation{$^6$Departament de Matem\`atiques, Universitat de les Illes Balears, E-07122 Palma, Spain.}

\begin{abstract}
In the last few years there is an unprecedented progress in the increase of the
power conversion efficiency of perovskite solar cells. Evidently, further advances of
the efficiency of these devices will depend
on the constraints imposed by the optical and electronic properties of their constituents.
Quite apparently that during the manufacturing process of  a solar cell,
 there is an inevitable variation in the thicknesses of various functional layers,
 which affects the optoelectronic characteristics of the final sample.
In this work a possible strategy of the analysis of the
solar cell performance is suggested, based on statistically averaging
procedure of experimental data.
 We present a case study, in
 which the optoelectronic properties of the meso-structured perovskite solar
 cell  (with a mesoporous TiO$_2$ layer) are analysed within  the method providing
a deeper understanding of the device operation.
This method enables an assessment of the overall quality of the device, pointing pathways towards
the maximum efficiency design of a perovskite  solar cell by  material properties tuning.
\end{abstract}
\maketitle

\begin{widetext}
\begin{center}
\end{center}
\end{widetext}

\section{INTRODUCTION}
\label{s1}
Nowadays organic-inorganic halide perovskites, as solar cell light-absorbers,
attract substantial attention. Solar cells with such photoactive layers are
considered as one of the most promising competitors to silicon-based photovoltaics,
reaching efficiencies of $25.8\%$ only in  two decades of development \cite{Min2021}.
Low-cost production of these materials with high power conversion efficiency (PCE)
pledge fast development of flexible solar cells \cite{Babu2020}.
Moreover, there is  challenging future in improvement of their performance in order
to reach their maximum efficiency (cf.\cite{MarEff}) by means:
engineering of perovskite composition \cite{Urzua-Leiva2020} or/and
perovskite solar cell (PSC) constituents \cite{fron,engen,Mar, Li2021, Zhang2021}; study and improvement
of perovskite stability \cite{Pereyra2021}; altering characteristics of optoelectronic behavior
of PSC constituents \cite{Bonnin-Ripoll2019} by incorporation, for example,
light trapping mechanisms \cite{Deng2019}; to name just a few.

 Evidently, for further efficiency improvements of PSCs, the analysis
of optical and electric properties of different layers in the  device architecture
is pivotal  for optimizing light-harvesting in the absorber, whilst
allowing coherent effects and parasitic absorption to be accounted in the stack design.
In addition, this analysis should be based on statistically reliable estimates of
noticeable dispersion of various sample parameters
during the manufacturing process.
Partially, some attempts in this direction have been done
in \cite{Haidari2019,Mishra2020,Husainat2020}, where various type of simulations of optoelectronic
properties of PSCs have been conducted recently.   It is worth noting that for the analysis of optical properties various
research groups (see, for example,
\cite{Hattab2021,Hussain2021,Yadav2022,Ahmad2022,Karthick2022}) use the software
SCAPS-1D \cite{scaps1dmanual} as a main tool. This code does not, however, consider interference,
scattering, or intermediate reflectors (see details in \cite{scaps1dmanual}),
thus preventing the comprehensive analysis of the antireflection layer's interference  and the
influence of thin films thicknesses.

Also, a broad range of electron (hole) transport materials
and back metal contacts have been considered to examine
the performance of lead-free PSCs in \cite{DeepthiJayan2021}.
It also became common to introduce a mesoporous TiO$_2$ layer between  TiO$_2$ electron
collecting contact and perovskite (the main absorber layer). This process decreases
defect density at the absorber boundary, while may introduce additional resistance.
Nevertheless, the effect of a mesoporous TiO$_2$ layer and its influence
on the PSC performance remains in infancy.
Here, we employ a ray tracing-based optoelectronic computational
model \cite{Bonnin-Ripoll2021} to reproduce the behaviour of a perovskite solar cell
with a mesoporous TiO$_2$ layer and characterize experimental PSC samples.
To this aim we further developed this
model, incorporating a few novel elements to treat properly
thin-film meso-structured perovskite solar cells.

\begin{figure*}
\centering
\begin{subfigure}[h]{0.56\textwidth}
       \includegraphics[width=0.56\textwidth,angle=90]{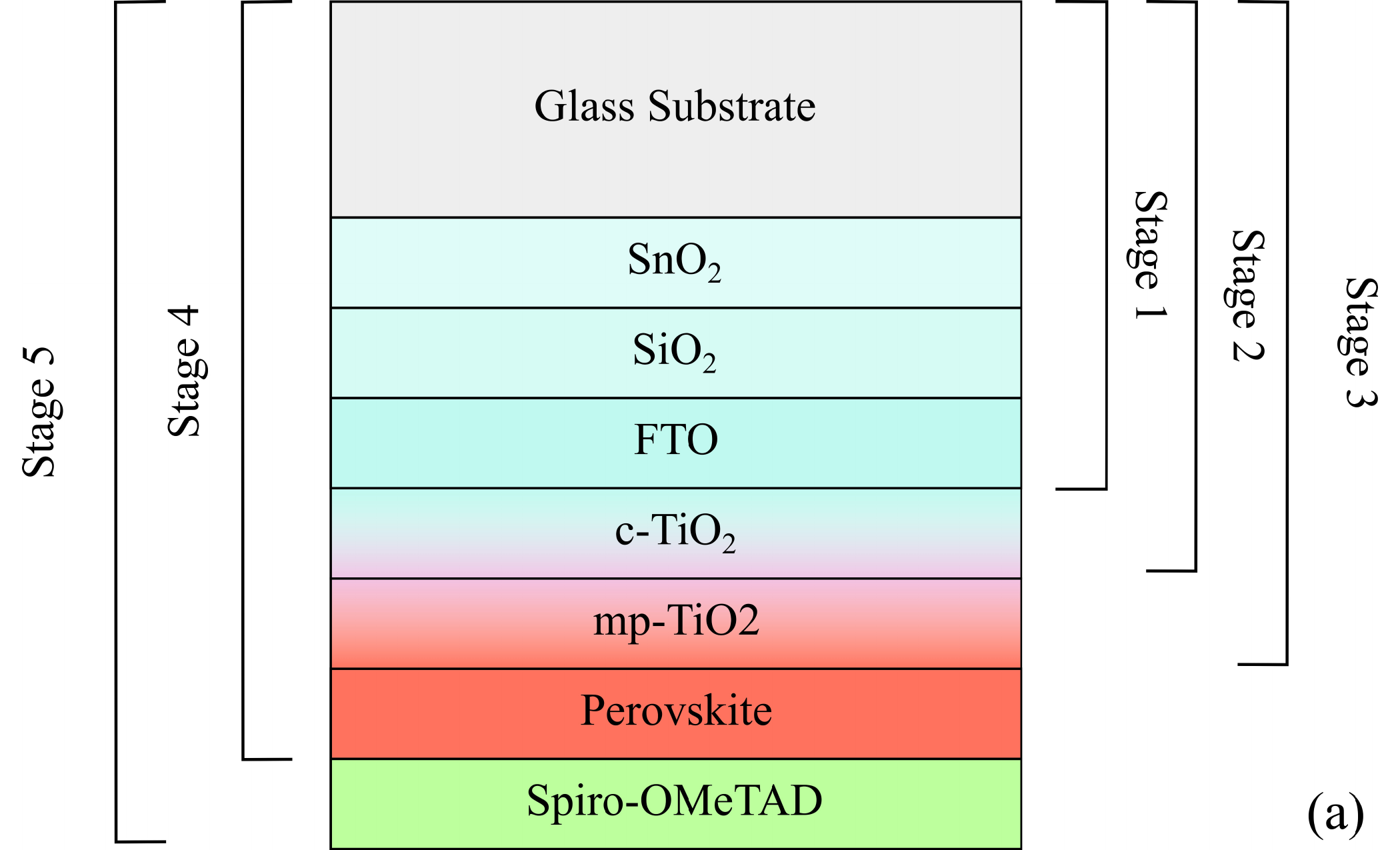}
\end{subfigure}
\begin{subfigure}[h]{0.33\textwidth}
       \includegraphics[width=\textwidth]{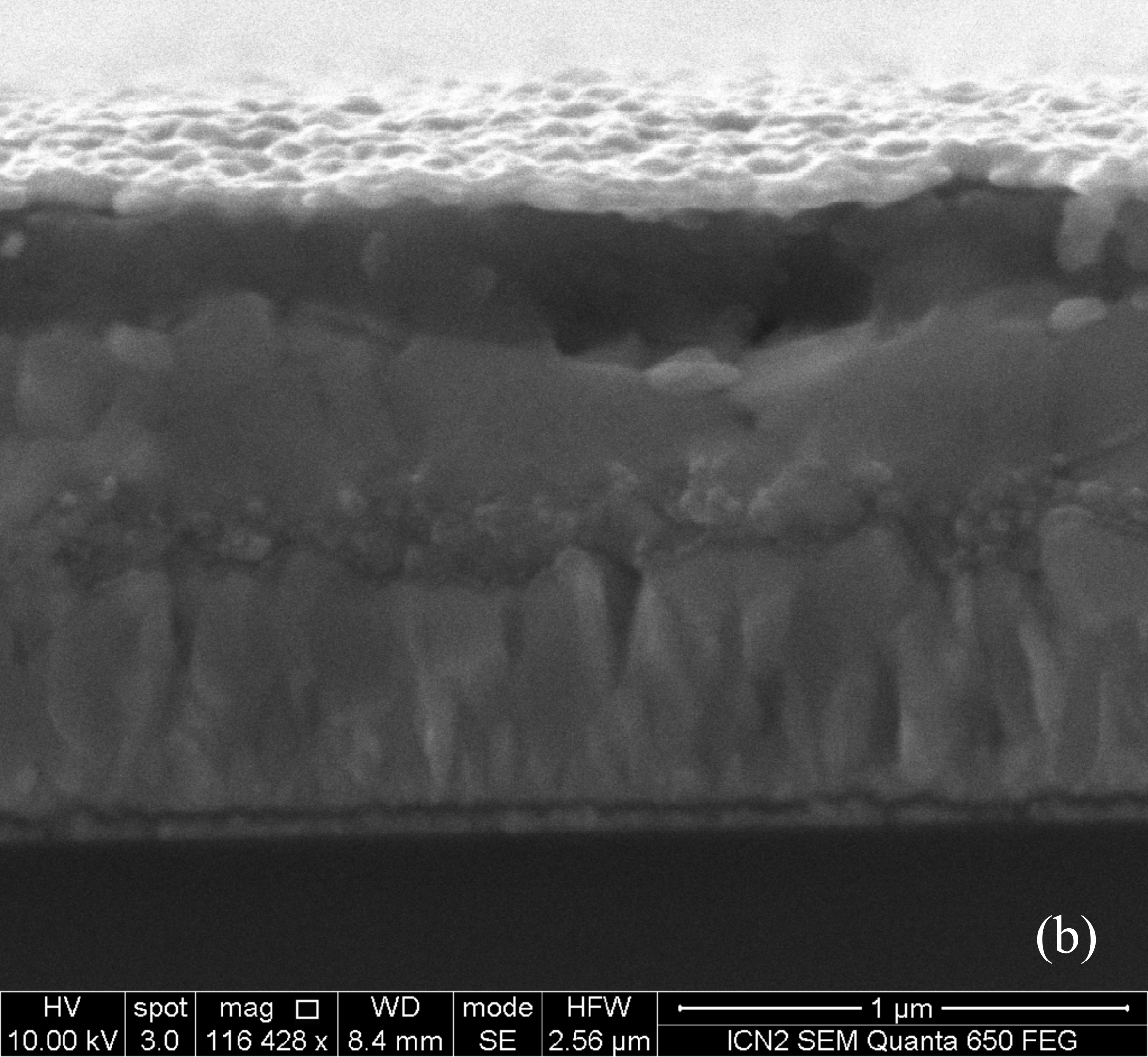}
\end{subfigure}
 \caption{(a) Structure of the layers present in each manufacturing stage.
 A total of five stages in which the optical behavior has been studied.
 (b) Typical cross-section image of the sample, obtained by means of the scanning electron microscope.}
 \label{fig:scheme}
\end{figure*}

The experimental device configuration (see Fig.\ref{fig:scheme})
is selected as the well established experimentally perovskite solar cell \cite{Xie2021}.
The measurements of optical transmittance have been performed in each stage
of the manufacturing process.
In this n-i-p junction device structure, apart the complex glass substrate,
a thin and compact TiO$_2$ (c-TiO$_2$) is manufactured
as an electron selective contact to facilitate the collection of photogenerated electrons from
perovskite absorber. Next, to improve the efficiency of carrier collection
a mesoporous TiO$_2$ (mp-TiO$_2$) is introduced to provide a suitable perovskite
deposition \cite{Liu2016}. In addition, the partial infiltration of halide perovskite
into the mp-TiO$_2$ layer facilitates higher quantum yield for photo-excited charge
separation. Spiro-OMeTAD  has been utilized in establishing a good p-contact
between the gold (Au) counter electrode  with
a perovskite absorber in our stack design.
As mentioned above, we will elaborate the model \cite{Bonnin-Ripoll2021}
and discuss in detail the effective medium  approach for analysis of
the optical behaviour of mp-TiO$_2$. Our approach is experimentally verified with
 measurements of the device reflection and transmission spectra.
These characteristics
have been measured by means of spectrophotometry techniques applied at each manufacturing stage.
Finally, we make use of our approach to evaluate the dependence of the PSC performance
on the properties of the mesoporous structure of the considered solar cell architecture,
based  on statistically averaged physical and geometrical parameters of the device.

\section{Methodology}
\label{s2}

\subsection{Device Characteristics}
\label{sub22}
Since electronic and optical measurements cannot be performed on the same PSC sample,
we use two batches of 24 solar cells, manufactured at the same conditions.
One batch is used for optical measurements with interruptions on each phase of the process,
while the second one is used for J-V characteristic measurements without
any interruptions on each phase of the process.

Considering the deposition of a material as a new phase of the process, we have
 performed the measurements of optical transmittance of each PSC layer in each stage
 of the manufacturing process (see details in Appendix \ref{sec:sup-data}).
 Stage 1 of the manufacturing process consists of the glass substrate, assuming
 the material as a multilayer system itself, formed by sodalime glass,
 SnO$_2$, SiO$_2$ and fluorine doped tin oxide (FTO). In the stage 2
 a layer of c-TiO$_2$  of the order of tens of nanometers is
 added by the spray pyrolysis technique. Next, a mp-TiO$_2$  layer
 is deposited in stage 3, followed by the subsequent deposition of a perovskite (stage 4)
 and Spiro-OMeTAD (stage 5), the latter three using the spin coating technique.

Next, using a solar simulator, the J-V characteristics of the PSCs in operation
a second batch of $24$ solar cells
has been measured. The result of
this measurements provides the average value of the short-circuit current
$J_{sc}=20.07$ mA/cm$^2$ with a standard deviation  $0.99$ mA/cm$^2$ at irradiance $1000$ W/m$^2$.
In this study we will focus on reproducing the behaviour of the reverse scan of the J-V
characteristics. The scanning process conditions are discussed in \cite{Xie2021}.

\subsection{Optoelectronic Model}
\label{sub24}
To analyse the amount of transmitted, absorbed
and reflected light in each of the materials present in the PSCs,
we employ ray tracing simulations with the aid of the ray trace OTSun python
package \cite{Cardona2020}.
It is a Monte Carlo ray tracing where the optics is implemented
through the Fresnel optics equations in their most general
form, complemented by the transfer-matrix method (TMM) to account
for the interference phenomenon \cite{StevenByrnes,byrnes2020multilayer}.
This software allows us to perform the simulations by loading a geometry file designed
with the aid of FreeCAD \cite{FreeCAD}, an open-source software for the
parametric modeling of geometric figures in 3D.
\begin{table*}[ht!]
\begin{tabular}{ |p{2.1cm}||p{2cm}|p{2cm}|p{2cm}|p{2cm}|}
 \hline
 \multicolumn{5}{|c|}{Input parameters in our calculations (References)} \\
 \hline
     & c-TiO$_2$ & mp-TiO$_2$  & Perovskite &  Spiro-OMeTAD  \\
 \hline
 $\chi$ [eV]	& 4 \cite{Liu2014} & 4 \cite{Liu2014}& 3.75 \cite{hima} & 2.12 \cite{fei}\\
 E$_{\it g}$ [eV] & 3.05 \cite{hind} & 3.05 \cite{hind} & 1.66	\cite{Chen2019} & 3.1 \cite{xin} \\
 $m_{e}^{\star}$ & 1 & 1& 1 & 1   \\
 $m_{h}^{\star}$ & 1 & 1& 1 & 1  \\
 $N_{d/a}$[cm$^{-3}$]& $1\times 10^{18}$& variable & 0 & $1\times 10^{21}$ \\
$\mu$ [cm$^2$/Vs] & 0.006 \cite{woj} &0.006	\cite{woj} & variable  & 0.0001 \cite{sna}\\
$\varepsilon$ & 60 \cite{dutta} & 42.45**	 & 60 \cite{martin} & 3 \cite{gar} \\
$\ell_{\it D}$[nm] & 4 & 4 & variable  & 0.5 \\
  \hline
 \end{tabular}
 \caption{Characteristics of the considered semiconductors:
TiO$_2$, FAPbI$_3$ (Perovskite), Spiro-OMeTAD.
We use the following notations:
$\chi$ is an electron affinity;
 E$_{\it g}$ is an energy gap; $m_{e/h}^{\star}$ is an effective electron/hole mass;
$N_{d/a}$, is a donor/acceptor concentration;
 $\mu$ is a mobility; $\varepsilon$ is a permittivity;
 $\ell_{\it D}$ is a diffusion length.(**)
 The definition of the permittivity of the mp-TiO$_2$
 is discussed in Sec.\ref{sub26};
 see Eq.\ref{mix} and the accompanying discussion.}
 \label{tabparam2}
\end{table*}
The simulations require knowledge of the layer thicknesses as well as
the complex refractive indexes of the corresponding materials
as a function of wavelength ($\lambda$).
In particular, the complex refractive indexes of
the following materials have been taken from literature:
FTO \cite{Ching-Prado2018}, TiO$_2$ \cite{Siefke2016}, Spiro-OMeTAD \cite{Chen2015}.
In the case of SnO$_2$ and SiO$_2$ the data are taken from the material manufacturer itself.
According to the experimental setup \cite{Xie2021} the chemical structure
of the perovskite layer has been associated with  FaPbI$_3$.
In this case the complex refractive index data are taken from \cite{Chen2019}, where
the real and imaginary parts of the refractive index have been measured for a similar perovskite.
The physical characteristics of all layers that form the architecture of our device
are displayed in Table \ref{tabparam2}.

On the other hand, if the layer contains
a few constituents there is a need in some effective approach.
The refractive index of the glass has been calculated from
transmittance experimental measurements, while for the
mp-TiO$_2$ material we have developed the effective approach to determine
its complex refractive index (see below).

\subsection*{Glass complex refractive index}
Since the glass substrate consists of a few thin components, to determine
the glass optical properties itself the oxides films have to be removed.
To this aim, on the glass multilayer system the etching process  is done by
using Zn powder and HCl solution. As a result, the optical properties of the glass
substrate has been determined with the aid of the method \cite{Nichelatti2002},
based on the ideas discussed below.
If the slab of material has plane-parallel faces, and a material is absorbing
partially ($k^2\ll n$), the real part $n$ and the imaginary part $k$ of the refractive
index of the slab can be calculated from the intensity of reflectance $R(\lambda)\equiv R$
and transmittance  $T(\lambda)\equiv T$ spectra (for the sake of convenience we omit the symbol $\lambda$).
In particular, the following analytical formulas have been used:
\begin{equation}
\label{eqn:n}
    k=\frac{\lambda}{4\pi d}\ln\Bigr[\frac{R_F\cdot T}{R-R_F}\Bigr]\,,
\end{equation}

\begin{equation}
\label{eqn:k}
        n = \frac{1+R_F}{1-R_F} + \Bigr[ \frac{4R_F}{(1-R_F)^2}-k^2\Bigr]^{1/2}\,,
\end{equation}
where the slab geometrical thickness, $d$, is assumed as known, and

\begin{widetext}
\begin{equation}
\label{eqn:RF}
    R_F = (2+T^2-(1-R)^2-\{ [2+T^2-(1-R)^2]^2-4R(2-R)\}^{1/2})(2(2-R))^{-1}\,.
\end{equation}
\end{widetext}

\begin{figure*}[!htb]
\centering
\begin{subfigure}[h]{0.44\textwidth}
       \includegraphics[width=\textwidth]{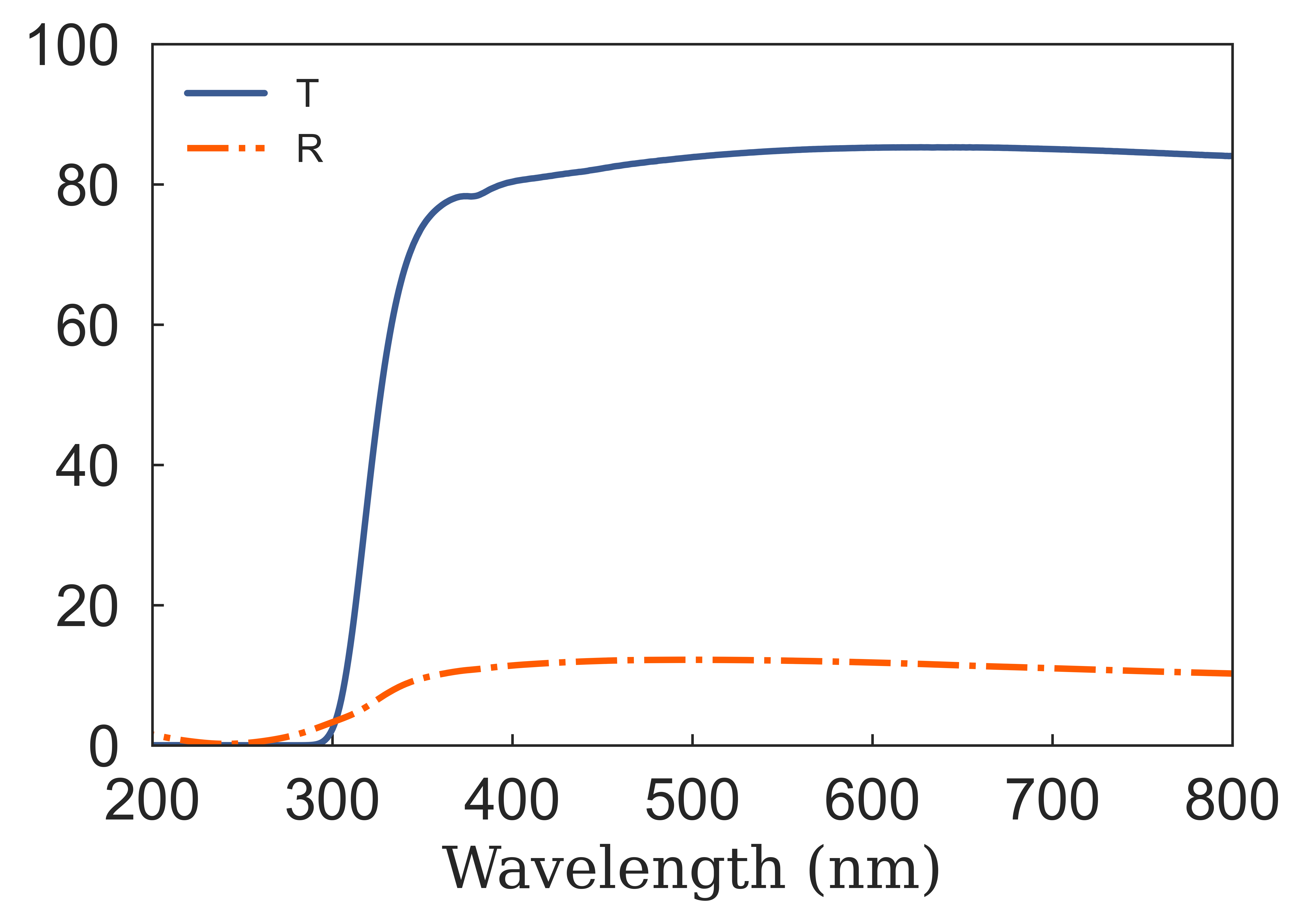}
    \end{subfigure}
\begin{subfigure}[h]{0.48\textwidth}
       \includegraphics[width=\textwidth]{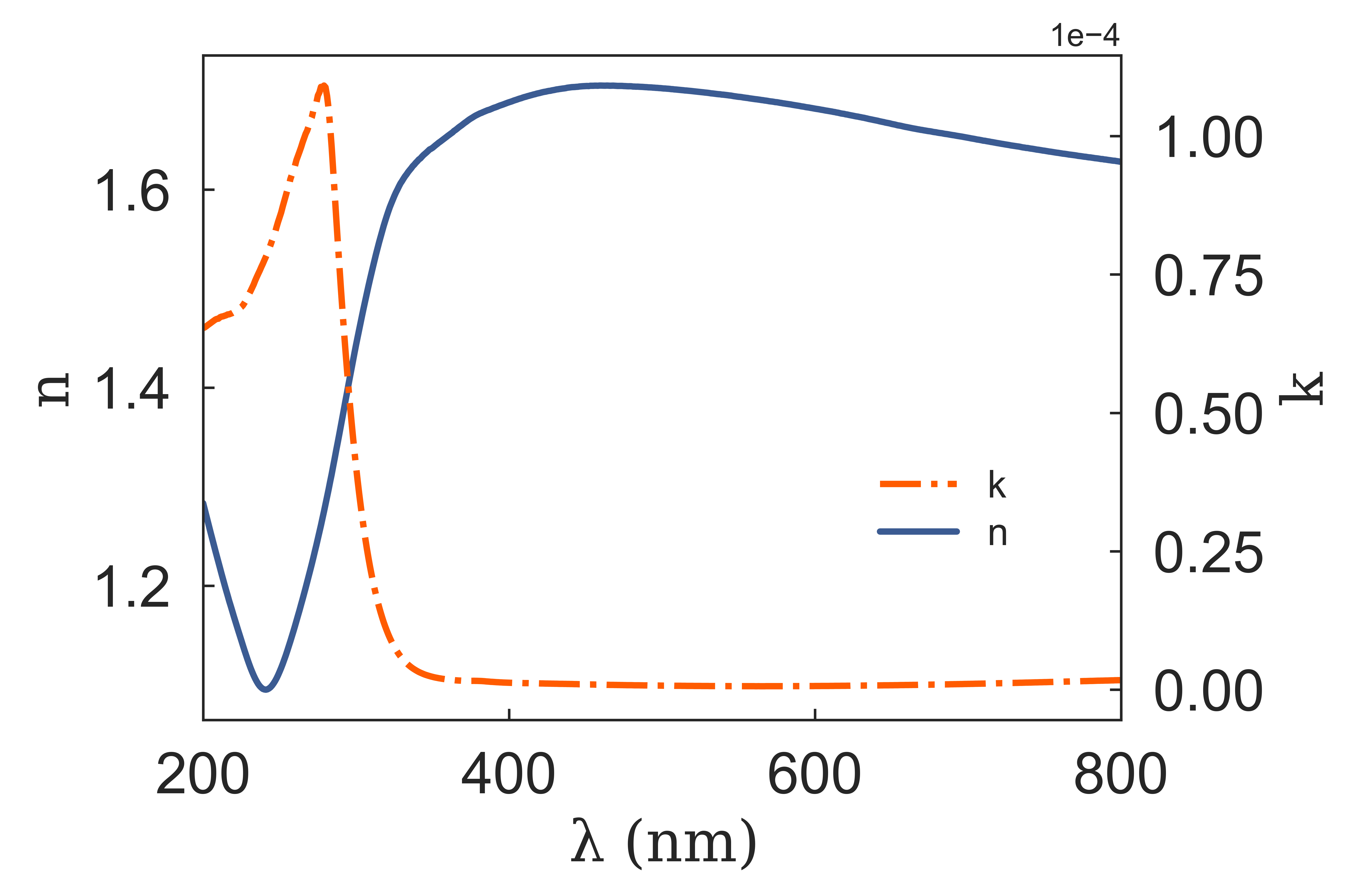}
    \end{subfigure}
\caption{Left panel: the results of transmittance and reflectance measurements of the
glass substrate as a function of the wavelength are shown.
Right panel: the real ($n$) and imaginary ($k$) components of the calculated complex refractive
index of the glass substrate as a function of the wavelength are displayed.}
\label{fig:TR_nk}
\end{figure*}

Experimental results for the transmittance $T(\lambda)$ and reflectance $R(\lambda)$ spectra,
and the complex refractive index, obtained by means of Eqs.(\ref{eqn:n})--(\ref{eqn:RF}),
are shown in Fig.\ref{fig:TR_nk}.

\subsection*{Mesoporous layer as an effective medium: an optical approach}

Mesoporous TiO$_2$ is one of the commonly used  electron transport materials (ETMs) in PSCs.
The mp-TiO$_2$ film serves not only as a scaffold for the perovskite layer
but also as a route for electron transport. However, in many simulations
 (see, for example, \cite{Kim2021,Widianto2021,Mehrabian2023,Hossain2023}) the effect of the porosity
of the  mesoporous TiO$_2$ layer on
optical and electronic properties meso-structured perovskite solar cells is ignored.

The presence of mesoporous thin films requires
a proper approach to determine its optical characterization.
Several effective medium approximations have been
suggested to describe the optical properties of heterogeneous thin films
by considering them as a homogeneous media with an effective refractive index,
along with effective absorption coefficient. Most of the models have been developed
for the effective refractive index omitting the absorption coefficient (see for a review \cite{Hutchinson2010}).
However, the absorption can affect the optoelectronic properties of ETM, and,
consequently, the properties of the whole system.

To consider the complex refractive index we apply the Volume Average Theory (VAT) model,
discussed among others in \cite{Hutchinson2010}.  It provides the effective complex
refractive index (subindex eff) of a material (see Eqs.(\ref{eqn:neff})--(\ref{eqn:B}))
composed of a continuous phase material matrix (subindex c) and a dispersed phase material
matrix (subindex d), with a porosity $p$:
\begin{equation}
\label{eqn:neff}
    n_{eff}^2=\frac{1}{2}(A+\sqrt{A^2+B^2})
\end{equation}
\begin{equation}
\label{eqn:keff}
    k_{eff}^2=\frac{1}{2}(-A+\sqrt{A^2+B^2})
\end{equation}
\begin{equation}
\label{eqn:A}
    A=p(n_d^2-k_d^2)+(1-p)(n_c^2-k_c^2)
\end{equation}
\begin{equation}
\label{eqn:B}
    B=2pn_dk_d+2(1-p)n_ck_c
\end{equation}
The properties of the mp-TiO$_2$ material can be characterized using an effective medium that
contains a mixture of the c-TiO$_2$ material and air, as suggested in 
\cite{Raoult2019, Raoult2022,Hernandez-Granados2019}.
In the latter paper, however, the effect of the extinction coefficient $k$ is missing.
Evidently, it may play important role, since the absorption in the mp-TiO$_2$ layer
depends on the degree of its porosity.
To find thickness configurations that match best the experimental results
we developed the special procedure to quantify the minimal deviation of
the theoretical estimations from the experimental values for the
transmittance and the short-circuit current 
(see Figs.\ref{fig:Jsc_acumulat},~\ref{fig:JscRMSE} and the accompanying text). 
Once the above analysis has been completed, to gain much more accurate results on
the optical properties of the PSC we use the OTSun python package.

\subsection*{Impact of porosity and Lambertian scattering}
\label{sub26}

The knowledge of the thickness configuration enables to us
to simulate the photogeneration rate for its application into
the electronic transport equations. For the sake of convenience
we recapitulate briefly the basic steps of this procedure
(see details in \cite{Bonnin-Ripoll2021}).

According to the Beer-Lambert law for a given
absorption coefficient $\alpha(\lambda)=4\pi k(\lambda)/\lambda$  of a material,
the loss of light intensity is a function of the path length  $\ell$ of the
light beam through a material:
\begin{equation}
\label{lambert}
N(\ell,\lambda)=N_0(0,\lambda)e^{-\alpha(\lambda)\mathsf{\ell}}\,.
\end{equation}
Here, $N_0 (0,\lambda)$ is the initial number of photons of
the incident ray on a material
per unit area, per wavelength, and per unit time (for each time when a ray impacts
onto a material according to the path trajectories
determined by the ray tracing simulation).
With the aid of OTSun and Eq.(\ref{lambert}), we calculate
the generation rate produced by a single ray
\begin{equation}
\label{t}
G(z)=\int_{0}^{\infty}\frac{\alpha(\lambda)N_0(0,\lambda)}{\cos\theta}
e^{-\alpha(\lambda)z/\cos\theta}d{\lambda}\,,
\end{equation}
where $z=\ell\cos\theta$ is the depth from a semiconductor
surface, and $\theta$ is the ray refracted angle.
The function $G(z)$ provides
the number of electrons per unit volume and per unit time, generated at each point in the
device due to the photon absorption. The total function $G(z)$ is the sum
of all rays (transmitted or reflected) passed at the point $z$.
As a result, obtained PSC thickness configuration and
the rate function $G(z)$ allow to calculate  the J-V characteristics by means of the transport
model \cite{Bonnin-Ripoll2021}. In this model there is a set of parameters that
are relatively well established either from literature or from  measurements.
Besides, there is another set that consists of some uncertain parameters.
 All of them are listed in Table \ref{tabparam2}, where uncertain parameters are named as ``variable".
 The latter ones are treated as fitting parameters to explain the measured data.

There are a few comments in order, however.
At this stage we have to elaborate the model \cite{Bonnin-Ripoll2021},
taking into account the presence of the mp-TiO$_2$ layer. Consequently, to carry out
the calculations within the transport model we have to define the permittivity and
the conductivity of the mp-TiO$_2$ layer $\sigma_{mp}=q\mu_{mp}N_{mp}$;
here $\mu_{mp}$ is a majority carriers mobility, and $N_{mp}$ is the doping concentration.

From the optical properties of different layers, we obtain the fact
that the mp-TiO$_2$ layer  includes air-filled holes (porosity).
As the concentration of these holes is small ($\sim 20\%$),
we apply the electrodynamic approach developed for weakly mixed systems
(see the textbook \cite{LL}).
In this case the permittivity of the mp-TiO$_2$ material is provided by
the expression for permittivity of the mixture $\varepsilon_{mix}$:
\begin{equation}
\label{mix}
\varepsilon_{mix}=\varepsilon_{TiO_2}+c
\frac{3(\varepsilon_a-\varepsilon_{TiO_2})\varepsilon_{TiO_2}}{\varepsilon_a+2\varepsilon_{TiO_2}}\,,
\end{equation}
where $\varepsilon_{TiO_2}$ and $\varepsilon_a$ are the permittivities of
TiO$_2$ and air, respectively; $c$ is the concentration of holes.
As for the conductivity, we make use of the fact obtained from
the optical properties fitting.
Due to small concentration of air-filled holes
in the mp-TiO$_2$ layer one
can suppose that  the conductivity in this layer should be reduced on 20\% as well.
 As a result, it seems reasonable to assume that
for the  donor concentration holds the relation
$N_{mp}=0.8\cdot N_c$, while the mobility of the carriers remains
the same $\mu_{mp}=\mu_c$, as in the c-TiO$_2$ layer (see Sec.\ref{s4})
To obtain the better agreement with the J-V characteristics of our
experimental device, we propose  to vary only the unknown
parameters $N_{mp}$, the mobility of carriers $\mu$  and their relaxation
time $\tau$ in the perovskite layer, keeping the other parameters fixed.

It is important to stress that  OTSun provides a more accurate form
of the photogeneration rate $G(z)$ compare to the TMM approach used
in Appendix \ref{sec:sup-data}, since non-specular reflection and multiple
reflection are considered. In particular, from images obtained
with the aid of the scanning electron microscope (see Fig.\ref{fig:SEM})
it can be seen that the gold layer (the top layer) has nano-sized roughness,
of the order of magnitude of its thickness ($\sim80$ nm).
With a high degree of certainty, it can be argued that these textures
provide a Lambertian reflection to the surface. Such  textures randomly
reflect the light that impacts on the gold layer, and, consequently, increase
the optical path in the perovskite, and  the photocurrent (cf.\cite{Bonnin-Ripoll2019}).
As a result, taking into account all facts above, we obtain
by means of Eq.(\ref{t}) that
$J_{sc}=20.15$ mA/cm$^2$ (see Fig.\ref{fig:JscRMSE}), i.e., it is
slightly higher than the TMM result and closer to the experimental value.

\begin{figure}[!htb]
\centering
 \includegraphics[scale=0.5]{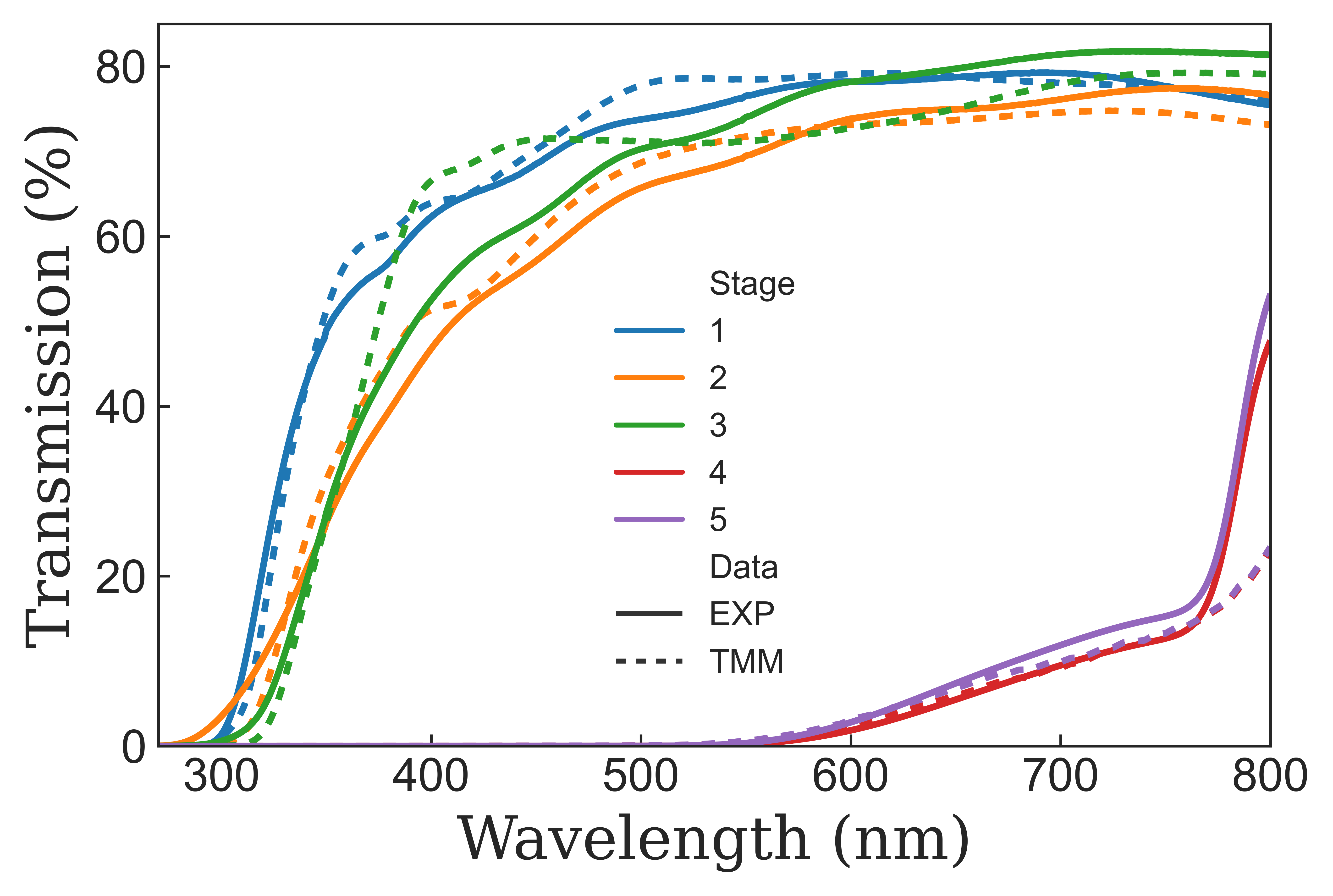}
\caption{Experimental (solid line) and theoretical
(dotted line) transmittance curves for the different manufacturing stages
of the PSCs under study. For the stage  identifications see also Fig.\ref{fig:scheme}}
\label{fig:optfit}
\end{figure}

\section{RESULTS AND DISCUSSION}
\label{s4}

Begin with, we recall the main result on the analysis of the layer thicknesses
(see details in Appendix \ref{sec:sup-data}).
The layer configuration chosen as the theoretical model of the PSCs studied
is (in nm): $11$ (SiO$_2$), $16$ (SnO$_2$), $565$ (FTO), $24$ (c-TiO$_2$),
$240$ (mp-TiO$_2$, porosity = $20\%$), $500$ (Perovskite), $250$ (Spiro-OMeTAD).
The result of this final disposition that produces the
best consistency between theoretical and experimental values for the short-circuit
current and the optical transmittance in each stage of the manufacturing process is displayed in
Fig.\ref{fig:optfit}.

There is a rather well agreement between  experimental and theoretical results,
except two cases: stage 3 and stage 4. The visible difference between the experimental and theoretical curves
is located in the wavelength region $380-500$ nm.
This result implies some inaccuracy in estimation of the mp-TiO$_2$ layer absorption coefficient. Theoretical result
exhibits the larger absorption for the perovskite FAPbI$_3$ (stage 4) in comparison with that of the experimental
sample in the wavelength range $770-800$ nm. This disagreement is due to the difference between
the experimental value of the perovskite absorption coefficient and that adopted from \cite{Chen2019}.
Since this difference is located at a small interval $760<\lambda<800$ nm  of the right boundary of
the considered wavelength region,
it may be safely ignored in the description of  the J-V characteristics discussed below.

The values of theoretical layer thicknesses and material parameters (see Table \ref{tabparam2}) are
used as the input parameters in the electronic transport model to fit the simulated and measured
J-V characteristics of the PSC. The fitting procedure is described in \cite{Bonnin-Ripoll2021}
but here we mention some peculiarities:
\begin{enumerate}
\item	The main source of difference between the simulated and the measured short-circuit
current has been related in \cite{Bonnin-Ripoll2021} to the reflection
properties of the PSC layers. In present paper the reflection properties are
already taken into account by our optical model. It results in the remarkable
accord between theoretical and experimental values of the short-circuit current (see below).

\begin{figure}[!htb]
\centering
 \includegraphics[scale=0.5]{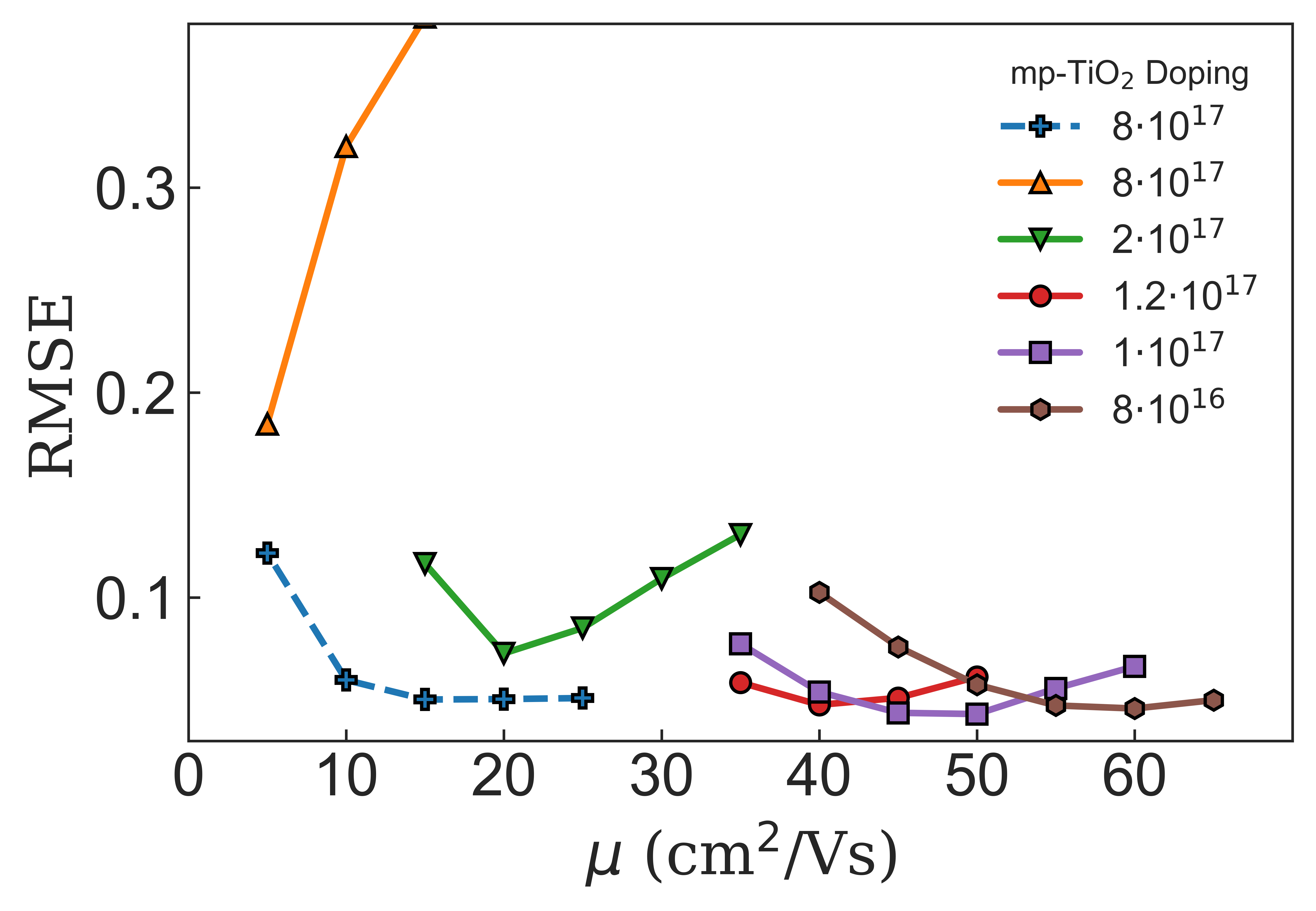}
 \caption{RMSE as a function of carrier mobility in perovskite
 for different doping concentration in the mp-TiO$_2$ layer.
 The solid curves are associated with the results of calculations
 without any additional resistances. The dotted line is associated
 with the results obtained by means of  the additional resistances (see details in the text).}
 \label{fig:mob}
 \end{figure}

 \item To fit the simulated and “measured averaged” open circuit voltage ($V_{oc}$)
 one needs to vary  the carrier lifetimes ($\tau$) in the perovskite \cite{Bonnin-Ripoll2021}.
 Once we vary the donor concentration $N_{mp}$ and the mobility  $\mu$, it is required
 to fit a $\tau$-value to reproduce the experimental value of $V_{oc}$.
 To control the accuracy of the fitting procedure of J-V characteristics,
 we introduce the root mean square error
 \begin{equation}
 \label{rm}
RMSE = \frac{1}{N}\sqrt{\sum_i
    [J_{sc}^{exp}(V_i)-J_{sc}^{th}(V_i)]^2}\,,
\end{equation}
where the number of steps $N$ is defined by the voltage range interval
$V=0-V_{oc}$ passed with a step size $\Delta V$.
 At each new value of the voltage $V_i=i\cdot \Delta V$, we solve the
 system of nonlinear transport equations (see Sec.2.4 in
 \cite{Bonnin-Ripoll2021}) to calculate the current and compare its
 value with the experimental data.
The results of calculations manifest the dependence of the RMSE
on values of  $N_{mp}$ and the mobility  $\mu$ (see Fig.\ref{fig:mob}).

 \item

\begin{figure*}[!htb]
 \centering
 \includegraphics[scale=0.46]{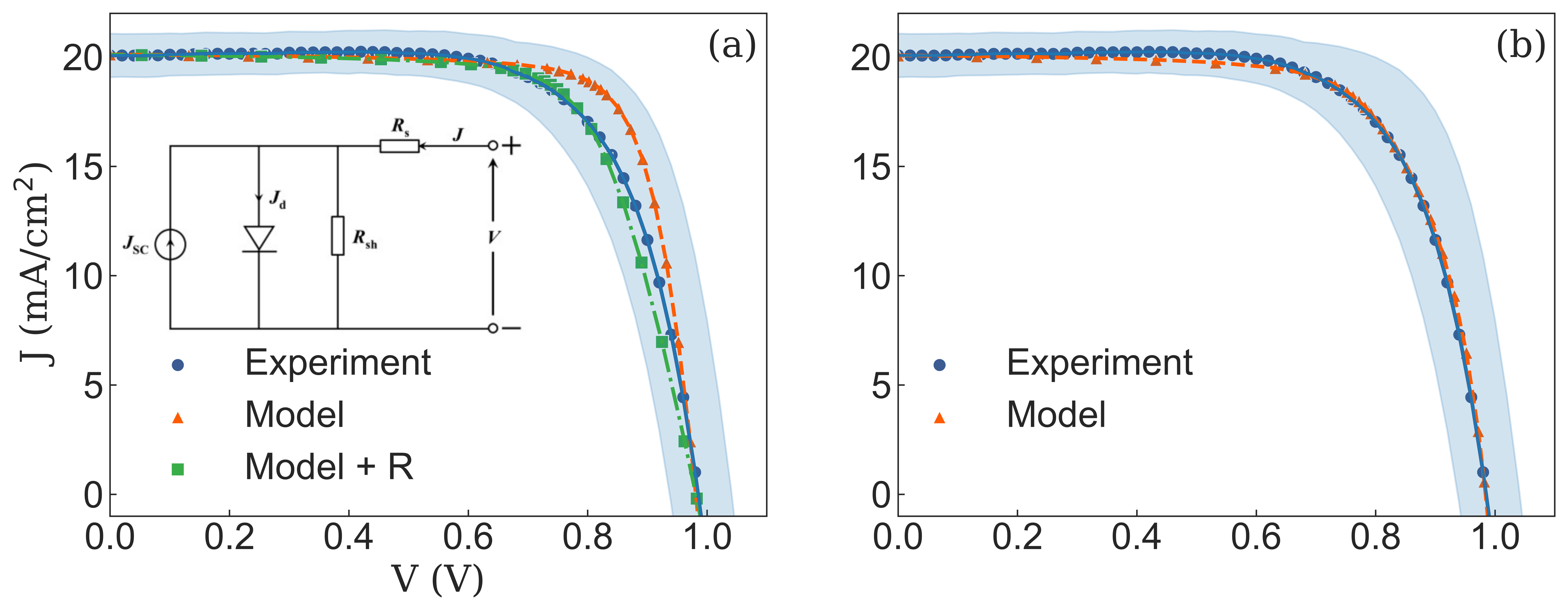}
\caption{J-V characteristics.
In panels (a) and (b) solid line connecting solid circles is
associated with the “measured averaged” values.
The panel (a): dashed (dot-dashed) line connecting solid triangles (squares) is associated with the
results obtained without (with) additional resistances at $20\%$ porosity of mp-TiO$_2$ layer.
The panel (b): dashed line connecting solid triangles is associated with the
results obtained without additional resistances (see for details the text).
Light blue band around all these curves marks the range
in which the
measured J-V characteristics of different samples lie.
The insert in the panel (a) displays the circuit layout of the generator.}
 \label{fig:JVC}
 \end{figure*}

We recall that the analysis of the transmittance
resulted in a degree of porosity of the order of 20\% for the mp-TiO$_2$ layer.
Evidently, the presence of the porosity in the mp-TiO$_2$ layer
should affect the donor concentration in this material in comparison
to that in the c-TiO$_2$ layer, which is $1\cdot 10^{18}$ cm$^{-3}$ (see Table \ref{tabparam2}).
Consequently, it is natural to suppose that
the donor concentration in the mp-TiO$_2$ layer will decrease
to the value $8\cdot 10^{17}$ cm$^{-3}$,
proportionally to the degree of its porosity.
Taking into account that the porosity affects the permittivity (see Eq.\ref{mix})
and the donor concentration of the mp-TiO$_2$ layer, we solve the transport
equations and reach the RMSE minimum (0.18 mA/cm$^2$) at $\mu=5$ cm$^2$/(V$\cdot$s)
(see Fig.\ref{fig:mob}; triangles connected by solid yellow line).
There is a visible difference between
the simulated and the “measured averaged” J-V characteristics (see Fig.\ref{fig:JVC}(a)).

A fairly common trick in the calculation of J-V characteristics
is to use the analogy of the PSC to a current generator in parallel
with a diode (see, for example, the analysis in \cite{Bonnin-Ripoll2019}).
Hereafter we follow this procedure, since
such an approach enables to us to consider the
parasitic resistance of the contacts and feeding conductors, R; the
leakage resistance, related to the local short-circuit of
electron/hole transport material layers, R$_{sh}$
(see the insert in Fig.\ref{fig:JVC}(a)).

Once all necessary additional resistances are generated,
we obtain a good agreement of our results with the experimental J-V characteristics.
This agreement (see Fig.\ref{fig:JVC}(a)) is reached
with the aid of the following values of the variable parameters:
$\mu$=15 cm$^2$/(V$\cdot$s) (see Fig.\ref{fig:mob}); R=4 Ohm$\cdot$cm$^2$;
R$_{sh}\rightarrow \infty$; $\tau=2.47$ ns that yields
the diffusion length $\ell_D=\sqrt{\mu k_BT\tau/|e|}=0.3$ $\mu$m.

It turns out, however, that we can obtain even better agreement
varying the donor concentration in the mp-TiO$_2$ layer.
The variation of the RMSE between simulated and
“measured averaged” J-V characteristics
yields the RMSE absolute minimum (see Fig.\ref{fig:mob})
 at the “optimal” mobility $\mu$=50 cm$^2$/(V$\cdot$s),
and the donor concentration $N_{mp}= 1\cdot 10^{17}$ cm$^{-3}$.
The optimal carrier lifetimes $\tau$=0.7 ns yields again
the diffusion length $\ell_D=0.3$ $\mu$m.
At these parameters we obtain a remarkable agreement between the simulated and
the “measured averaged” J-V characteristics, without any additional resistances
(see Fig.\ref{fig:JVC}(b)).
It seems that in the mp-TiO$_2$ layer pore holes  generate all necessary
additional resistance in the measured device.
Consequently, it appears that in the real device the pore dimension
occupying 20\% of the mp-TiO$_2$ layer
 is sufficient to affect optical properties of the PSC. It turns out, however, that only
 small part of the remaining TiO$_2$ take part
in the conductance due to large amount of defects on the pore boundaries.
\end{enumerate}

It is noteworthy, the parameters values obtained by means of our fitting procedure
are located within the parameter range considered in literature.
In particular, several studies suggest that the doping values of the mp-TiO$_2$ material are in
the range of $10^{16}$ to $10^{19}$ (cm$^{-3}$) \cite{Sellers2011}.
Moreover, the carrier mobility range measured for perovskite is in the range from
5 to 50 (cm$^2$/(V$\cdot$s))\cite{Herz2017}. The degree of the accuracy of our
analysis is summarized in Tables \ref{tab:results},\ref{tab:results_param}.
It demonstrates that the short-circuit current, the open circuit voltage,
the fill factor and the power conversion efficiency for both simulated and experimental
PSC are in a remarkable agreement.

\begin{table}[!ht]
\centering
\begin{tabular}{|c|c|c|c|c|}
\hline
&\textbf{$J_{sc}$ (mA/cm$^2$)} &\textbf{$V_{oc}$ (V)}&\textbf{FF}&\textbf{PCE(\%)}\\
\hline
\textbf{Experiment} & $20.072$ & $0.985$ & $0.695$ & $13.735$\\
\textbf{Model (a)} & $20.120$ & $0.981$ & $0.772$ & $15.233$\\
\textbf{Model (a) + R} & $20.101$ & $0.981$ & $0.705$ & $13.903$\\
\textbf{Model (b)} & $20.076$ & $0.984$ & $0.703$ & $13.882$\\

\hline
\end{tabular}
\caption{Main PSC parameters for the “measured averaged” device and simulated results
obtained by means of our approach. Model (a) is associated with results displayed on the
the panel Fig.\ref{fig:JVC}(a), without any resistance;
the corresponding parameters are presented in the first column
of Table \ref{tab:results_param}. Model (a) + R  is associated with results displayed on the
the panel Fig.\ref{fig:JVC}(a), with resistance;
the corresponding parameters are presented in the second column
of Table \ref{tab:results_param}.
Model (b) is associated with results displayed on the
the panel Fig.\ref{fig:JVC}(b);
the corresponding parameters are presented in the third column
of Table \ref{tab:results_param}.}
\label{tab:results}
\end{table}

\begin{table}[!ht]
\centering
\begin{tabular}{|c|c|c|c|}
\hline
\textbf{Parameters} & \textbf{Model (a)} & \textbf{Model (a) + R} & \textbf{Model (b)} \\
\hline
\textbf{$N_{mp}$ (cm$^{-3}$)} & $8 \cdot 10^{17}$ & $8 \cdot 10^{17}$ & $1 \cdot 10^{17}$\\
\textbf{$\mu$ (cm$^2$/(V$\cdot$s))} & $15$ & $15$ & $50$\\
\textbf{$\tau$ (ns)} & $2.47$ & $2.47$ & $0.7$\\
\textbf{R (Ohm$\cdot$cm$^2$)} & $-$ & $4$ & $-$\\
\textbf{R$_{sh}$ (Ohm$\cdot$cm$^2$)} & $-$ & $\infty$ & $-$\\
\hline
\end{tabular}
\caption{Summary of the parameters used to represent Fig.\ref{fig:JVC} and Table \ref{tab:results}.}
\label{tab:results_param}
\end{table}

\section{CONCLUSIONS}
\label{s5}

In summary we have successfully utilized the optoelectronic model based on
experimentally obtained optical features for planar meso-structured perovskite solar cells.
Our optoelectronic model incorporates ray tracing simulations, the transfer-matrix method and
the carrier transport equations. The ray tracing allows the exploration of any geometry, since the solar collector
device is designed by means of the FreeCAD software. The  combination of the ray tracing with the transfer-matrix method
 allows to perform the optical simulations considering Fresnel optics, internal reflections of the multilayer system,
 scattering produced by nanoroughness layers, and effective layers with some peculiarities (such as porous materials).

Large number of transmittance-reflectance measurements on the several stages
of manufacturing process have been used to fix the uncertainties of
the PSC layers thicknesses and the mesoporous TiO$_2$ layer porosity.
The same number of measurements have been done to define
the average J-V characteristics of  experimental devices.
The obtained results within the optoelectronic model
have been further used to analyse J-V characteristics of the manufactured PSCs
by means of the optimization procedure based on the multiple-use of the solution
transport equations \cite{Bonnin-Ripoll2021}. The developed model demonstrates  good agreement
with both optical and electrical measurements
leading to extraction of the layers thicknesses, the porosity degree and the permittivity
of the mesoporous TiO$_2$ material, the perovskite carriers mobility and lifetimes.
In addition, we have found that even small degree of the porosity can affect the performance
of the experimental devices, reducing essentially the effective donor concentration
(i.e., the conductance) of the mp-TiO$_2$ layer in comparison to that in the c-TiO$_2$ layer.

Our approach allows the study of devices with different architectures.
Reversing the order of the materials in the solar cell structure does
not affect the development of the optoelectronic simulations.
Thus our work provides a useful platform
for device optimization towards increasing PCE as well as appropriate tools for the analysis of
various optoelectronic properties of meso-structured perovskite solar cells.

\section*{Acknowledgements}
\label{s6}
This work is part of the project TED2021-132758B-I00, funded by MCIN/ AEI /10.13039/501100011033/ and the European Union ``Next Generation EU''/PRTR. It has also been co-funded by the predoctoral contracts call of the Vice-presidency and Innovation, Research and Tourism Department of the Government of the Balearic Islands and the European Social Fund (ESF) [Grant FPI/2144/2018]. Part of this work is under Physics Ph.D. programme for F. Bonnín-Ripoll of the Universitat de les Illes Balears (UIB, Spain). We thank Dr José F. González Morey (Serveis Cientificotècnics, UIB) for his technical assistance. This work is partially supported by the Russian Science Foundation (grant no. 23-19-00884).
We thank the Spanish State Research Agency for the grant Self-Power (PID2019-104272RB-C54 / AEI / 10.13039/501100011033), the ProperPhotoMile Project (PCI2020-112185) and the OrgEnergy Excelence Network (CTQ2016-81911- REDT), and to the Agència de Gestió d’Ajuts Universitaris i de Recerca (AGAUR) for the support to the consolidated Catalonia research group 2021 SGR 01617 and the Xarxa d’R+D+I Energy for Society (XRE4S). Part of this work is under Materials Science Ph.D. Degree for K.T. and the Chemistry Ph.D. programme for C.P. of the Universitat Autònoma de Barcelona (UAB, Spain). ICN2 is supported by the Severo Ochoa program from Spanish MINECO (grant no. SEV-2017-0706) and is funded by the CERCA Programme/Generalitat de Catalunya.”

\appendix
\section{Supplementary data}
\label{sec:sup-data}

\begin{figure}[!htb]
\centering
\includegraphics[scale=0.2]{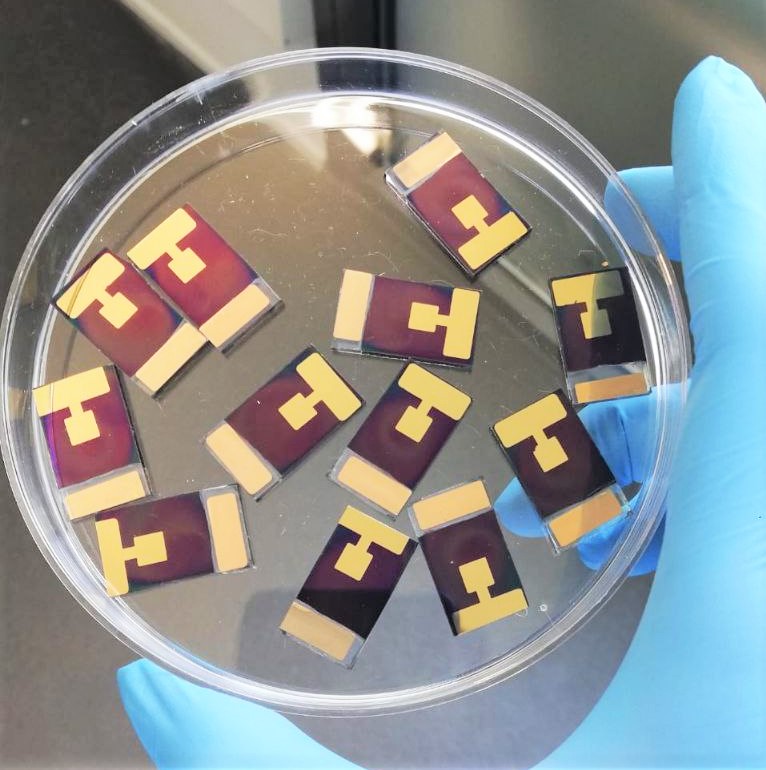}
\caption{Image of the perovskite solar cell samples at the
final stage of the manufacturing process.}
\label{fig:pict}
\end{figure}
\subsection*{Device Characteristics}
\label{sec:dev-ch}
The device fabrication and the experimental work, presented
in this paper, have been carried out at the facilities of
the Institut Català de Nanociència i Nanotecnologia (ICN2)
in collaboration with the Nanomaterials for Photovoltaics Energy research group.
The perovskite solar cells  have been fabricated with
the FTO/c-TiO$_2$/mp-TiO$_2$/Perovskite/Spiro-OMeTAD/Au
configuration using multi-cationic perovskite.
Each of these materials has its own synthesis and deposition process;
all fabrication details are discussed in \cite{Xie2021}.
Example of final samples is shown in Fig.\ref{fig:pict}.

\begin{figure}[!htb]
\centering
\begin{subfigure}[h]{0.45\textwidth}
      \includegraphics[width=\textwidth]{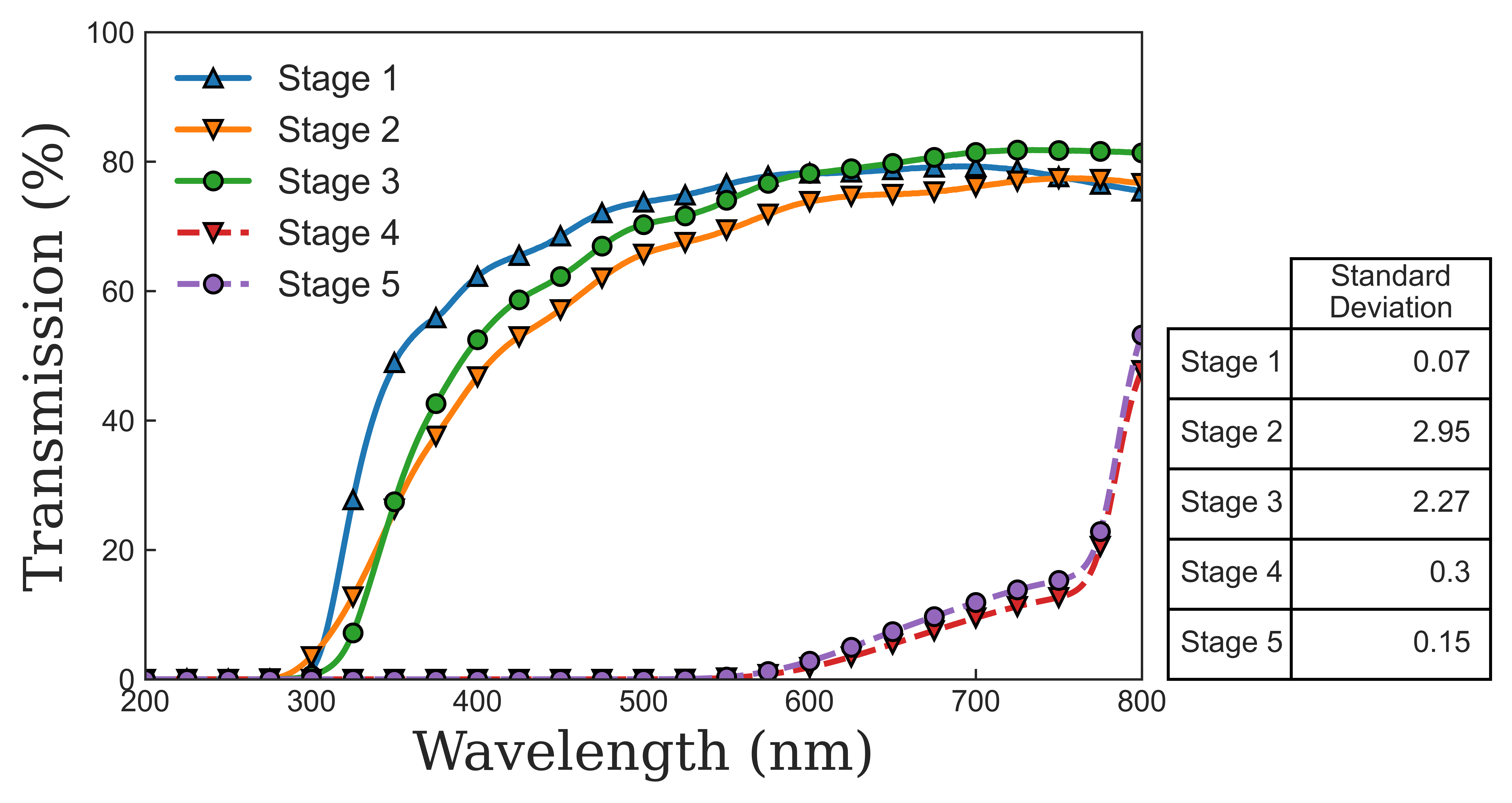}
\end{subfigure}
\caption{Mean values and standard deviations of the optical
transmittance measurements for each manufacturing stages of the perovskite
solar cell. Standard deviations have been calculated at each wavelength
for six experimental samples (see also the main text).}
\label{fig:T_stages}
\end{figure}

\subsection*{Transmittance}
\label{sec: trn-data}
Transmittance measurements have been done by means of the UV-Vis CARY 4000
spectrophotometer at normal incidence, using the solid sample holder accessory.
The incident radiation consists of $200-800$ nm wavelength range of the optical spectrum.
The results of the mean value of the optical transmittance measurements
for each stage  as a function of the wavelength are shown in
Fig.\ref{fig:T_stages}.
Our measurements make manifest that it is enough to fabricate six samples
at each stage to obtain a reliable result for the mean value.
Additionaly, we indicate the standard deviation from the mean value for each curve, corresponding
to each stage. The largest deviations are obtained for the c-TiO$_2$ and mp-TiO$_2$
depositions (stages 2 and 3, respectively). Noteworthy, up to now there is no any available
criterium in literature regarding definition of the effective border line
between c-TiO$_2$ and mp-TiO$_2$ depositions. In fact, one of our goals is
to propose such the effective methodology.

\begin{figure}[!htb]
\centering
\includegraphics[scale=0.49]{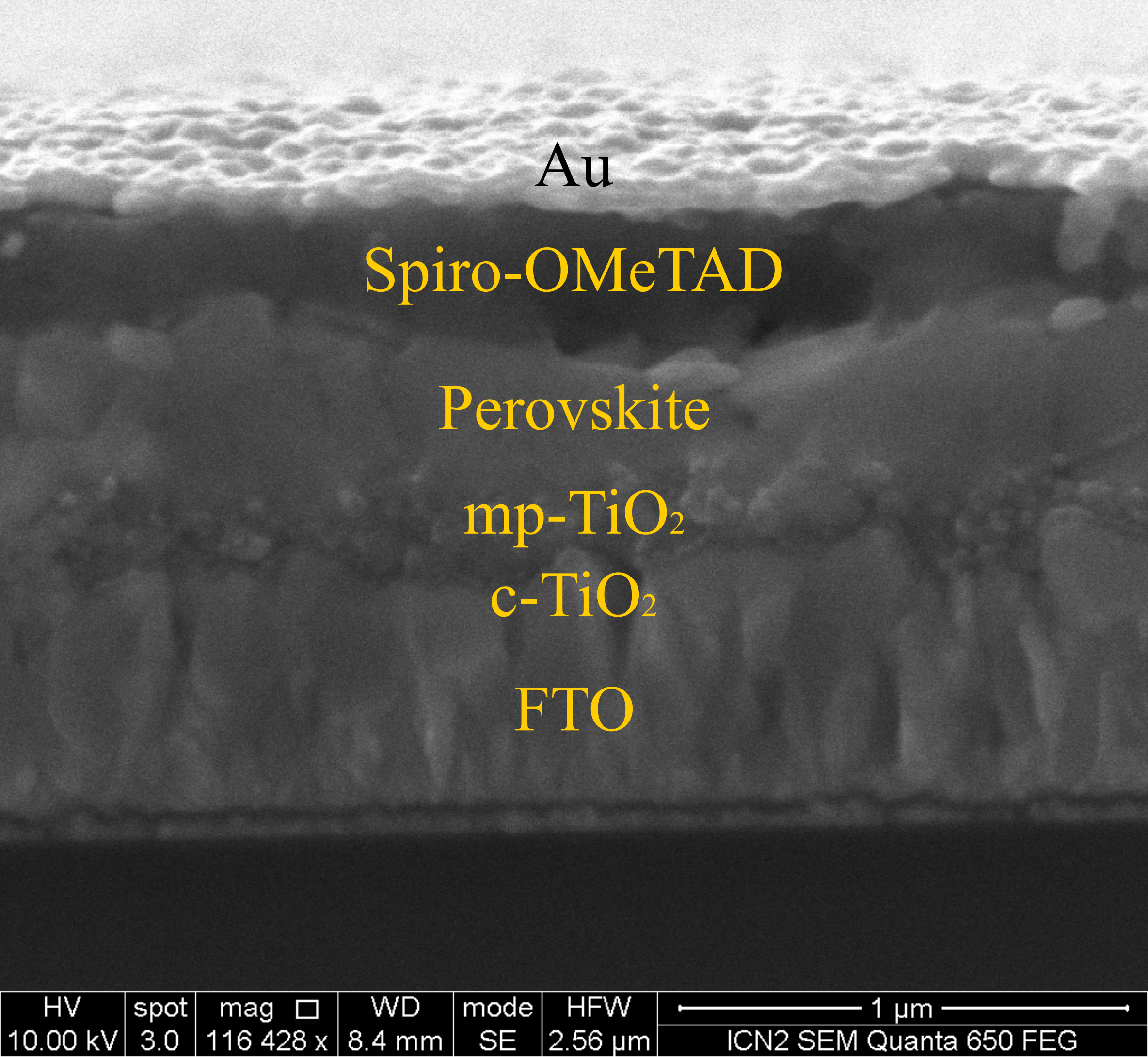}
\caption{Cross-section image of the samples, obtained by means of the scanning electron microscope.
It is quite visible the degree of the gold layer roughness. The extracted approximate thickness of each layer is
presented in Table \ref{tab:SEM}.}
 \label{fig:SEM}
\end{figure}

\subsection*{Layer Thicknesses}
\label{sec: thick-data}

The approximate thickness of each material layers has been estimated using scanning
electron microscopy (SEM) techniques (see Fig.\ref{fig:SEM}).
These  images  are essential for modelling the PSC, since each layer thickness assesment
is a key parameter in the following model calculations.
The heterogeneity observed in the layer thicknesses is due to the experimental process itself,
subject to variations in the initial conditions, the experience and skill of the technician,
the correct stoichiometric balance of the compounds, etc.
The less accuracy for the estimation of the c-TiO$_2$ layer thickness is due
to its extension diapason in the range of a few tens of nanometers.

\begin{table}[!ht]
\centering
\begin{tabular}{|c|c|}
\hline
\textbf{Material} &\textbf{Thicknesses (nm)}\\
\hline
SiO$_2$ & 10-30\\
SnO$_2$ & 10-30\\
FTO & 500-600\\
c-TiO$_2$ & 10-30\\
mp-TiO$_2$ & 180-300\\
Perovskite & 400-550\\
Spiro-OMeTAD & 200-250\\
\hline
\end{tabular}
\caption{Thickness range of the material layers present in the PSC samples
measured by scanning electron microscopy.}
\label{tab:SEM}
\end{table}

Note that the substrate is formed by four different materials.
The thickness of each component in the compound is provided by
the manufacturer itself: $2.2$ mm of glass, $25$ nm of SnO$_2$ and
SiO$_2$, and $540$ nm of FTO.
Even so, in the images (see Fig.\ref{fig:SEM}) it can be seen that the FTO is not
completely homogeneous. Consequently, we consider a certain range of
thicknesses, in which the calculations for the characterization are carried out.
From the images obtained by SEM, the thickness ranges considered for the materials
in the successive calculations, are shown in Table \ref{tab:SEM}.

\subsection*{Detailed elaboration of thickness configuration}
\label{sec:det-thick}

In order to fit theoretical and experimental transmittances, a concrete thickness
configuration of the PSC structure is needed. Since scanning electronic microscope
images show that the thicknesses of the layers is not completely homogeneous,
we must consider a range between which each layer thickness lies.
Before to resolve this problem we have to analyse the optical properties of each layer.

One of the most common theoretical tools used to analyse the optical properties of
complex solar systems with a high accuracy is the so called Monte-Carlo
ray tracing simulation. It consists of a set of techniques that determine
the path of light through matter in a three-dimensional environment
with computer simulations \cite{Bos17}.  We implement in
our approach the Monte-Carlo ray tracing technique, adding the
transfer-matrix method (TMM) to characterize the optical response of
the PSC with the aid of the ray trace OTSun python
package \cite{Cardona2018,OTSunWebApp,Tutorial2}.
In order to carry out simulations in a simpler and faster way,
first, we use the TMM only to determine the thicknesses configuration.
Based on the results discussed hereafter,
the OTSun will be used (see the main text) as the subsequent step where more
accurate calculations are needed. Consequently, using the ranges shown in Table \ref{tab:SEM},
each possible configuration is considered to provide the
transmittance with the aid of the TMM calculation,
to obtain those that are closer to the experimental curves.
To reach our goal we introduce the root mean square error
(RMSE) of each thickness configuration for a given material thickness
\begin{equation}
\label{eqn:RMSE}
    RMSE = \frac{1}{N}\sqrt{\sum_{\lambda=300}^{\lambda=800}
    [T^{exp}(\lambda)-T^{th}(\lambda)]^2}\,.
\end{equation}
 Here, $N$ is the total number of wavelength steps;
the step is equal to $1$ nm in all cases; $T^{exp}$($T^{th}$) is the experimental
(theoretical) optical transmittance. Hereafter, it is convenient to introduce the range
limits of the thickness variation for each layer as [a,b] (in nm).

First,  we carry out the calculations for the stage 1.
To this aim we consider the following thicknesses: [10,30] for SnO$_2$ and SiO$_2$;
[500, 600]  for FTO. As a result, we choose those configurations that lie within 1\% of the lowest RMSE.

Next, we proceed to the stage 2 (repeating similar calculations), using  the range limits
obtained from the stage 1: [10,20] for SnO$_2$ and SiO$_2$;
[550,600] for FTO; [10,30] for c-TiO$_2$.
These calculations reduce the starting range limits
to [10,16], [14,20], [550,600], [22,30] for SnO$_2$, SiO$_2$, FTO, c-TiO$_2$, respectively.

\begin{figure*}[!htb]
\centering
\includegraphics[scale=0.45]{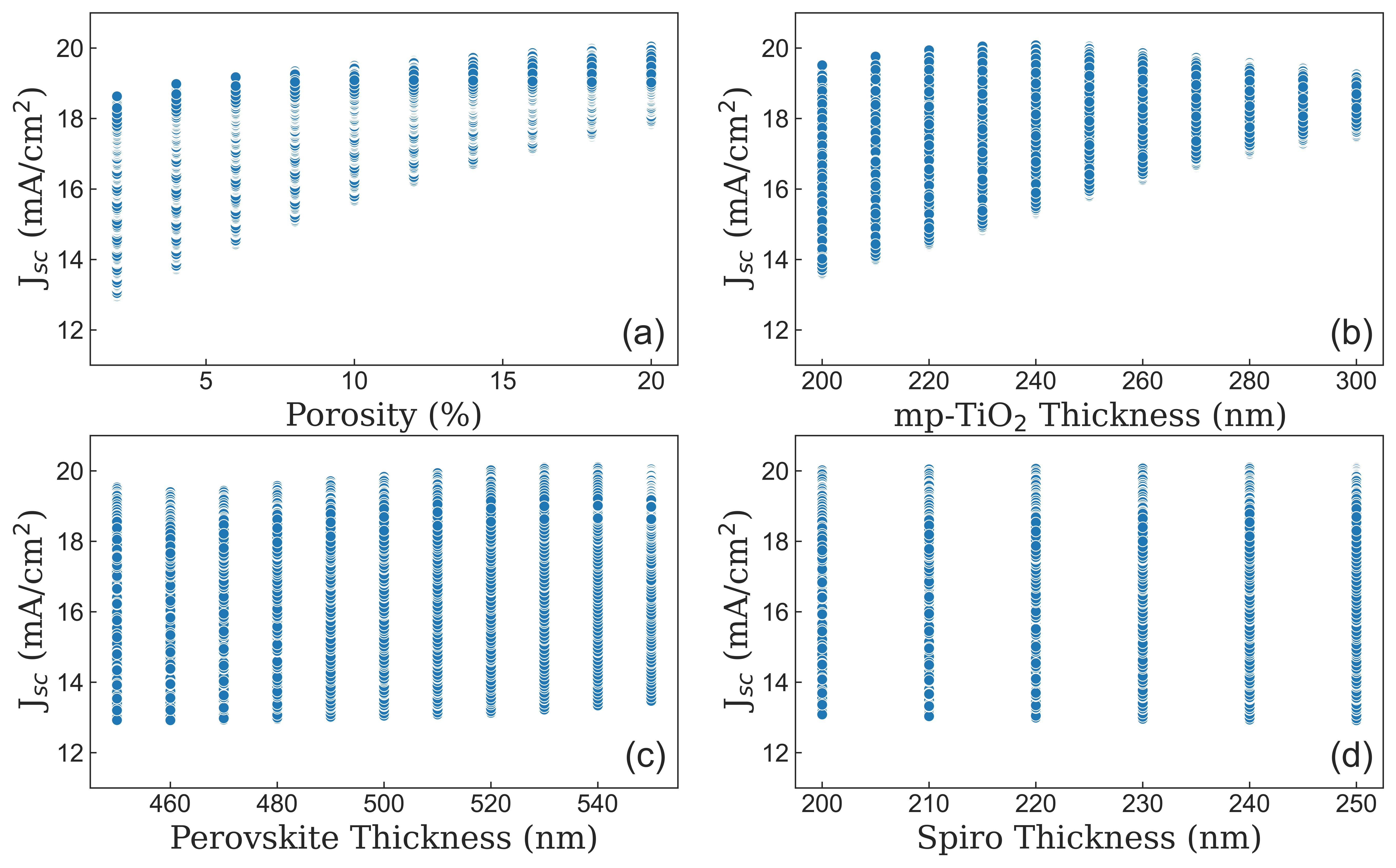}
\caption{The short-circuit current $J_{sc}$ as a function of mp-TiO$_2$ porosity (a),
mp-TiO$_2$ thickness (b), perovskite thickness (c) and Spiro-OMeTAD thickness (d).}
\label{fig:Jsc_acumulat}
\end{figure*}

Afterwards, we move to the stage 3, where both the thickness and the porosity of the mp-TiO$_2$
are added to the system. An initial sweep is made considering the previous ranges plus the mp-TiO$_2$ [180,300]
layer and its porosity considered between 1\% and 50\%. Considering the configurations that
lie within 1\% of the lowest RMSE, the material thicknesses are varied in the following
intervals: [10,12] for SnO$_2$; [14,18] for SiO$_2$; [555,580] for FTO; [23,25] for c-TiO$_2$;
[200,300] for mp-TiO$_2$, and its porosity reduced to the 10\%--20\% range.

From our previous experience (see details in \cite{Bonnin-Ripoll2019,Bonnin-Ripoll2021})
it follows that small changes in layer thicknesses (in order of a few nanometers) at
the stage 2 do not affect the optical transmittance of the device. Therefore, we
choose the mean values 11, 16, 565, and 24 nm (associated with SnO$_2$, SiO$_2$,
FTO, c-TiO$_2$ layer thickness, respectively), which are defined within the ranges
imposed by the stage 3. By fixing these values, we also reduce significantly
the calculation time, enabling us to explore the mesoporous material and
its subsequent influence on the properties of the absorbing layer.

We recall that the stage 4 includes the perovskite layer considered in the [400,550] interval,
while keeping the variability of the porosity and thickness of the mp-TiO$_2$ layer as mentioned above.
Since there is a complex interplay of the transmittance through the mp-TiO$_2$ layer and its effect
on the transmittance through the perovskite, we consider the ranges of three parameters simultaneously:
i) the mp-TiO$_2$ layer thickness; ii) its porosity; iii) the perovskite
thickness within 5\% of the lowest RMSE selection. As a result, we obtain that:
the mp-TiO$_2$ porosity oscillates between 10\% and 20\%, while its thickness
between 260 and 300 nm; the perovskite layer thickness oscillates between 450 and 500 nm.
In the stage 5 the Spiro-OMeTAD layer is added with a thickness defined in [200,250] nm interval.
Our calculations yield that the configurations within  5\% of the lowest RMSE lie in the following interval (in nm):
[270,300] for the mp-TiO$_2$ layer; the porosity is defined between 15\% and 20\%; [450,470] for
the perovskite layer;  [200, 250] for the Spiro-OMeTAD layer.

Notice that the configurations that fit better the optical behaviour of the experimental samples
are not necessarily those that fit better the electronic properties
of the experimental samples. Consequently, our analysis is supplemented by one
more parameter: the short-circuit current $J_{sc}$.
Therefore, considering the ranges of parameters for each material obtained under
the previous procedure and making them less restrictive, we  proceed to the
calculations of the short-circuit current $J_{sc}$, which will involve
adding the gold metal layer to the system. Again, we have to vary  the discussed
parameters within the above limits (see below).
Taking this fact into account, the maximum possible photocurrent
is calculated in the assumption that all absorbed photons generate
electron and holes and contribute to the current \cite{gar}.
In an efficient PSC this current is closely related to the short-circuit current, $J_{sc}$.
As a result, to reach a consistently between optical and electronic properties we calculate the
short-circuit current for every configuration, taking into account
that the mean value of the measured current $J_{sc}=20.07$ mA/cm$^2$ (see details in the main text).

We recall that all the effects regarding the radiation-matter interaction mentioned
above, are needed to elucidate the number of absorbed photons inside active materials.
To this aim the photogeneration rate $G$ of the creation of pairs of electrons/holes should be
considered a position-dependent function within the material thickness.
To this aim we employ  the method proposed by Ball et al. \cite{M.Ball2015}
and the TMM library \cite{StevenByrnes}
to calculate the short-circuit current $J_{sc}$ as:
 \begin{equation}
\label{eqn:Jsc}
    J_{sc} = q\int G(\lambda,z)d\lambda dz\,,
\end{equation}
where $q$ is the electron charge, and $z$ is the depth from the semiconductor surface.
\begin{figure}[!htb]
\centering
\includegraphics[scale=0.5]{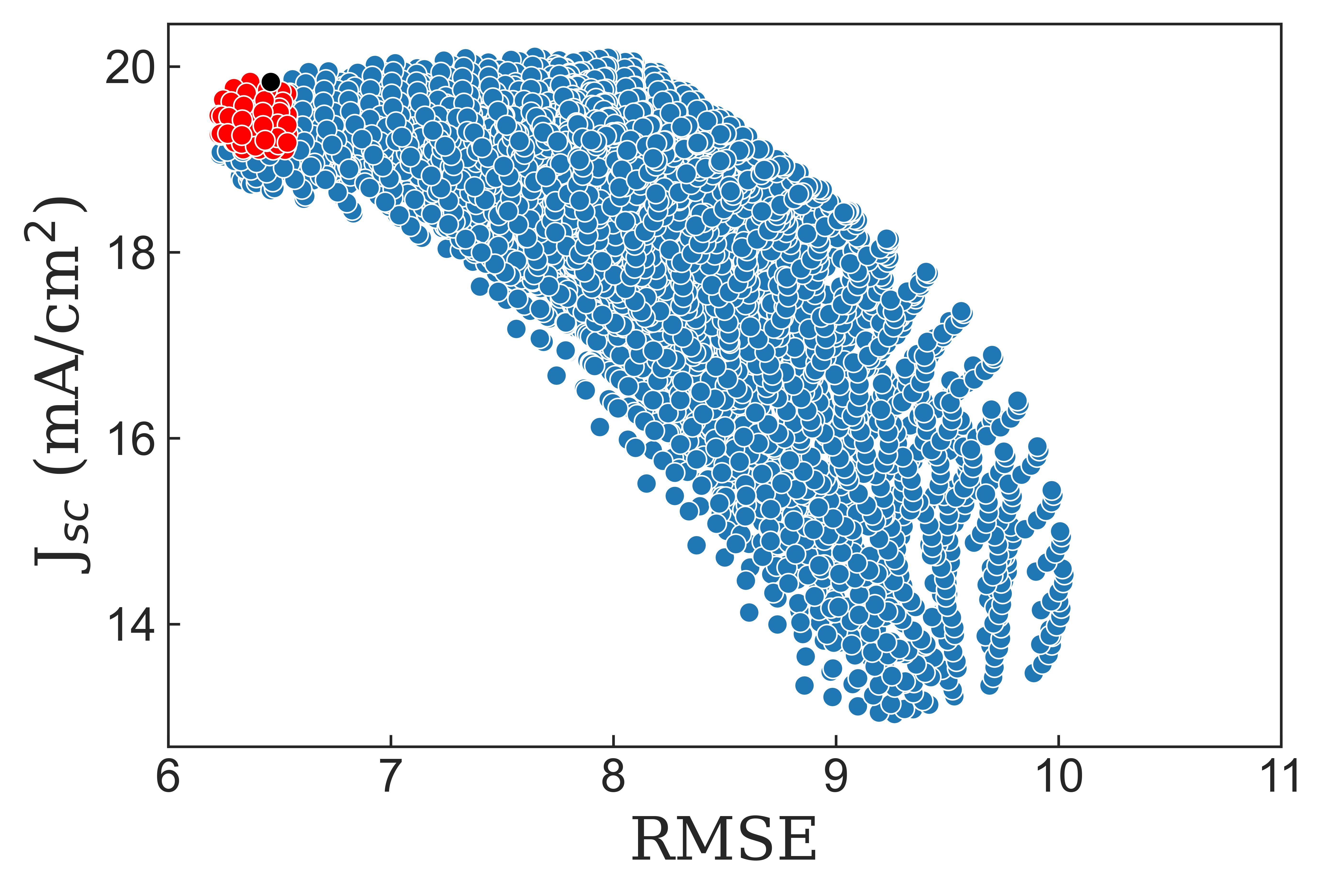}
\caption{The short-circuit $J_{sc}$ as a function of the RMSE, calculated by fitting
the experimental and theoretical transmittance of stage 5. The optimal
configurations (red solid points)
are located in the range of $\pm 5\%$ deviation from the experimental value
of the short-circuit current. Black point corresponds to the selected case, used for
comparison of experimental and theoretical J-V characteristics
(see main text).}
\label{fig:JscRMSE}
\end{figure}
Turning to the calculation details of the discussed procedure, first, we fix the
multilayer glass thicknesses at $11$, $16$, $565$ (nm) for SnO$_2$, SiO$_2$, FTO, respectively; and $24$ nm for c-TiO$_2$.
Second, it is observed that the photocurrent is very sensitive to the thickness and porosity of the mp-TiO$_2$ layer.
Indeed, it should be noted that at stages $2$ and $3$ that correspond to
the deposition of the c-TiO$_2$ and the mp-TiO$_2$ layers,
respectively,  a standard deviation $10$ times higher relative to those for
other stages (see Fig.\ref{fig:T_stages}).
Including the $J_{sc}$ calculations to the RMSE result obtained from the transmittance fitting,
we weakend the boundaries found above, in particularly, for the porosity and for the mp-TiO$_2$
layer thickness. Consequently, we vary the porosity and thickness of the mp-TiO$_2$ layer
(from 2\% to 20\%, from 200 to 300 nm, respectively); the perovskite thickness (from 450 to 550 nm);
the Spiro-OMeTAD thickness  (from 200 to 250 nm).
As a result, we do calculations
by varying the following parameters:
\begin{enumerate}
\item porosity takes 10 possible values;
\item mp-TiO$_2$ thickness takes 11 possible values;
\item perovskite thickness takes 11 possible values;
\item Spiro-OMeTAD thickness takes 6 possible values,
\end{enumerate}
i.e. we have a set of $10\times11\times11\times6=7260$ combinations
determined by the parameters variations.
 Altogether the chosen parameters
provide the basis for the calculation of the set  $J_{sc}$ values (see Fig.\ref{fig:Jsc_acumulat}).

Among all the possible configurations, we have selected those that hold the following criteria:
i) be $\leq |5|\%$  with respect to the experimental value of the short-circuit current $J_{sc}$;
ii) be $\leq 5\%$ of the lowest RMSE (red region in Fig.\ref{fig:JscRMSE}).
 From this selection, the configuration with the highest $J_{sc}$ has been chosen denoted by the black point on Fig.\ref{fig:JscRMSE}.
The results for the short-circuit current $J_{sc}$ demonstrate:
i) its less sensitivity to the variation of the perovskite and Spiro-OMeTAD thicknesses;
ii) while the close correspondence to the experimental value of
the short-circuit current $J_{sc}$ takes place
at the thickness $240$ nm and the porosity value = $20\%$  of the mp-TiO$_2$ layer.

Summarising, according to our analysis
the theoretical configuration, chosen to model the experimental samples, consists of
the following layer thicknesses (in nm): $11$ (SiO$_2$), $16$ (SnO$_2$),
$565$ (FTO), $24$ (c-TiO$_2$), $240$ (mp-TiO$_2$,
the porosity = $20\%$), $500$ (Perovskite), $250$ (Spiro-OMeTAD).
This configuration is obtained  as a result of
optimization of the optical transmittance of the device configuration
at the value of the short-circuit current
$J_{sc}=19.83$ mA/cm$^2$.

\bibliographystyle{apsrev4-2}
\bibliography{Paper3_bib}

\begin{thebibliography}{60}%
\makeatletter
\providecommand \@ifxundefined [1]{%
 \@ifx{#1\undefined}
}%
\providecommand \@ifnum [1]{%
 \ifnum #1\expandafter \@firstoftwo
 \else \expandafter \@secondoftwo
 \fi
}%
\providecommand \@ifx [1]{%
 \ifx #1\expandafter \@firstoftwo
 \else \expandafter \@secondoftwo
 \fi
}%
\providecommand \natexlab [1]{#1}%
\providecommand \enquote  [1]{``#1''}%
\providecommand \bibnamefont  [1]{#1}%
\providecommand \bibfnamefont [1]{#1}%
\providecommand \citenamefont [1]{#1}%
\providecommand \href@noop [0]{\@secondoftwo}%
\providecommand \href [0]{\begingroup \@sanitize@url \@href}%
\providecommand \@href[1]{\@@startlink{#1}\@@href}%
\providecommand \@@href[1]{\endgroup#1\@@endlink}%
\providecommand \@sanitize@url [0]{\catcode `\\12\catcode `\$12\catcode
  `\&12\catcode `\#12\catcode `\^12\catcode `\_12\catcode `\%12\relax}%
\providecommand \@@startlink[1]{}%
\providecommand \@@endlink[0]{}%
\providecommand \url  [0]{\begingroup\@sanitize@url \@url }%
\providecommand \@url [1]{\endgroup\@href {#1}{\urlprefix }}%
\providecommand \urlprefix  [0]{URL }%
\providecommand \Eprint [0]{\href }%
\providecommand \doibase [0]{https://doi.org/}%
\providecommand \selectlanguage [0]{\@gobble}%
\providecommand \bibinfo  [0]{\@secondoftwo}%
\providecommand \bibfield  [0]{\@secondoftwo}%
\providecommand \translation [1]{[#1]}%
\providecommand \BibitemOpen [0]{}%
\providecommand \bibitemStop [0]{}%
\providecommand \bibitemNoStop [0]{.\EOS\space}%
\providecommand \EOS [0]{\spacefactor3000\relax}%
\providecommand \BibitemShut  [1]{\csname bibitem#1\endcsname}%
\let\auto@bib@innerbib\@empty
\bibitem [{\citenamefont {Min}\ \emph {et~al.}(2021)\citenamefont {Min},
  \citenamefont {Lee}, \citenamefont {Kim}, \citenamefont {Kim}, \citenamefont
  {Lee}, \citenamefont {Kim}, \citenamefont {Paik}, \citenamefont {Kim},
  \citenamefont {Kim}, \citenamefont {Kim}, \citenamefont {Shin},\ and\
  \citenamefont {{Il Seok}}}]{Min2021}%
  \BibitemOpen
  \bibfield  {author} {\bibinfo {author} {\bibfnamefont {H.}~\bibnamefont
  {Min}}, \bibinfo {author} {\bibfnamefont {D.~Y.}\ \bibnamefont {Lee}},
  \bibinfo {author} {\bibfnamefont {J.}~\bibnamefont {Kim}}, \bibinfo {author}
  {\bibfnamefont {G.}~\bibnamefont {Kim}}, \bibinfo {author} {\bibfnamefont
  {K.~S.}\ \bibnamefont {Lee}}, \bibinfo {author} {\bibfnamefont
  {J.}~\bibnamefont {Kim}}, \bibinfo {author} {\bibfnamefont {M.~J.}\
  \bibnamefont {Paik}}, \bibinfo {author} {\bibfnamefont {Y.~K.}\ \bibnamefont
  {Kim}}, \bibinfo {author} {\bibfnamefont {K.~S.}\ \bibnamefont {Kim}},
  \bibinfo {author} {\bibfnamefont {M.~G.}\ \bibnamefont {Kim}}, \bibinfo
  {author} {\bibfnamefont {T.~J.}\ \bibnamefont {Shin}},\ and\ \bibinfo
  {author} {\bibfnamefont {S.}~\bibnamefont {{Il Seok}}},\ }\href
  {https://doi.org/10.1038/s41586-021-03964-8} {\bibfield  {journal} {\bibinfo
  {journal} {Nature 2021 598:7881}\ }\textbf {\bibinfo {volume} {598}},\
  \bibinfo {pages} {444} (\bibinfo {year} {2021})}\BibitemShut {NoStop}%
\bibitem [{\citenamefont {Babu}\ \emph {et~al.}(2020)\citenamefont {Babu},
  \citenamefont {{Fuentes Pineda}}, \citenamefont {Ahmad}, \citenamefont
  {Alvarez}, \citenamefont {Castriotta}, \citenamefont {{Di Carlo}},
  \citenamefont {Fabregat-Santiago},\ and\ \citenamefont
  {Wojciechowski}}]{Babu2020}%
  \BibitemOpen
  \bibfield  {author} {\bibinfo {author} {\bibfnamefont {V.}~\bibnamefont
  {Babu}}, \bibinfo {author} {\bibfnamefont {R.}~\bibnamefont {{Fuentes
  Pineda}}}, \bibinfo {author} {\bibfnamefont {T.}~\bibnamefont {Ahmad}},
  \bibinfo {author} {\bibfnamefont {A.~O.}\ \bibnamefont {Alvarez}}, \bibinfo
  {author} {\bibfnamefont {L.~A.}\ \bibnamefont {Castriotta}}, \bibinfo
  {author} {\bibfnamefont {A.}~\bibnamefont {{Di Carlo}}}, \bibinfo {author}
  {\bibfnamefont {F.}~\bibnamefont {Fabregat-Santiago}},\ and\ \bibinfo
  {author} {\bibfnamefont {K.}~\bibnamefont {Wojciechowski}},\ }\href
  {https://doi.org/https://doi.org/10.1021/acsaem.0c00702} {\bibfield
  {journal} {\bibinfo  {journal} {ACS Applied Energy Materials}\ }\textbf
  {\bibinfo {volume} {3}},\ \bibinfo {pages} {5126} (\bibinfo {year}
  {2020})}\BibitemShut {NoStop}%
\bibitem [{\citenamefont {Martynov}\ \emph {et~al.}(2017)\citenamefont
  {Martynov}, \citenamefont {Nazmitdinov}, \citenamefont {Moià-Pol},
  \citenamefont {Gladyshev}, \citenamefont {Tameev}, \citenamefont {Vannikov},\
  and\ \citenamefont {Pudlak}}]{MarEff}%
  \BibitemOpen
  \bibfield  {author} {\bibinfo {author} {\bibfnamefont {Y.~B.}\ \bibnamefont
  {Martynov}}, \bibinfo {author} {\bibfnamefont {R.~G.}\ \bibnamefont
  {Nazmitdinov}}, \bibinfo {author} {\bibfnamefont {A.}~\bibnamefont
  {Moià-Pol}}, \bibinfo {author} {\bibfnamefont {P.~P.}\ \bibnamefont
  {Gladyshev}}, \bibinfo {author} {\bibfnamefont {A.~R.}\ \bibnamefont
  {Tameev}}, \bibinfo {author} {\bibfnamefont {A.~V.}\ \bibnamefont
  {Vannikov}},\ and\ \bibinfo {author} {\bibfnamefont {M.}~\bibnamefont
  {Pudlak}},\ }\href {https://doi.org/10.1039/C7CP03892E} {\bibfield  {journal}
  {\bibinfo  {journal} {Phys. Chem. Chem. Phys.}\ }\textbf {\bibinfo {volume}
  {19}},\ \bibinfo {pages} {19916} (\bibinfo {year} {2017})}\BibitemShut
  {NoStop}%
\bibitem [{\citenamefont {Urz{\'{u}}a-Leiva}\ \emph {et~al.}(2020)\citenamefont
  {Urz{\'{u}}a-Leiva}, \citenamefont {{Narymany Shandy}}, \citenamefont {Xie},
  \citenamefont {Lira-Cant{\'{u}}},\ and\ \citenamefont
  {C{\'{a}}rdenas-Jir{\'{o}}n}}]{Urzua-Leiva2020}%
  \BibitemOpen
  \bibfield  {author} {\bibinfo {author} {\bibfnamefont {R.}~\bibnamefont
  {Urz{\'{u}}a-Leiva}}, \bibinfo {author} {\bibfnamefont {A.}~\bibnamefont
  {{Narymany Shandy}}}, \bibinfo {author} {\bibfnamefont {H.}~\bibnamefont
  {Xie}}, \bibinfo {author} {\bibfnamefont {M.}~\bibnamefont
  {Lira-Cant{\'{u}}}},\ and\ \bibinfo {author} {\bibfnamefont {G.}~\bibnamefont
  {C{\'{a}}rdenas-Jir{\'{o}}n}},\ }\href {https://doi.org/10.1039/D0NJ02748K}
  {\bibfield  {journal} {\bibinfo  {journal} {New journal of chemistry}\
  }\textbf {\bibinfo {volume} {44}},\ \bibinfo {pages} {14642} (\bibinfo {year}
  {2020})}\BibitemShut {NoStop}%
\bibitem [{\citenamefont {Frontera}\ \emph {et~al.}(2019)\citenamefont
  {Frontera}, \citenamefont {Martynov}, \citenamefont {Nazmitdinov},\ and\
  \citenamefont {Moià-Pol}}]{fron}%
  \BibitemOpen
  \bibfield  {author} {\bibinfo {author} {\bibfnamefont {A.}~\bibnamefont
  {Frontera}}, \bibinfo {author} {\bibfnamefont {Y.}~\bibnamefont {Martynov}},
  \bibinfo {author} {\bibfnamefont {R.~G.}\ \bibnamefont {Nazmitdinov}},\ and\
  \bibinfo {author} {\bibfnamefont {A.}~\bibnamefont {Moià-Pol}},\ }in\
  \href@noop {} {\emph {\bibinfo {booktitle} {{Perovskite Solar Cells:
  Properties, application and efficiency}}}}\ (\bibinfo  {publisher} {Nova
  Science Publishers/ Inc.},\ \bibinfo {year} {2019})\ Chap.~\bibinfo {chapter}
  {3}, pp.\ \bibinfo {pages} {117--174}\BibitemShut {NoStop}%
\bibitem [{\citenamefont {Agresti}\ \emph {et~al.}(2019)\citenamefont
  {Agresti}, \citenamefont {Pazniak}, \citenamefont {Pescetelli}, \citenamefont
  {Di~Vito}, \citenamefont {Rossi}, \citenamefont {Pecchia}, \citenamefont
  {Auf~der Maur}, \citenamefont {Liedl}, \citenamefont {Larciprete},
  \citenamefont {Kuznetsov}, \citenamefont {Saranin},\ and\ \citenamefont
  {Di~Carlo}}]{engen}%
  \BibitemOpen
  \bibfield  {author} {\bibinfo {author} {\bibfnamefont {A.}~\bibnamefont
  {Agresti}}, \bibinfo {author} {\bibfnamefont {A.}~\bibnamefont {Pazniak}},
  \bibinfo {author} {\bibfnamefont {S.}~\bibnamefont {Pescetelli}}, \bibinfo
  {author} {\bibfnamefont {A.}~\bibnamefont {Di~Vito}}, \bibinfo {author}
  {\bibfnamefont {D.}~\bibnamefont {Rossi}}, \bibinfo {author} {\bibfnamefont
  {A.}~\bibnamefont {Pecchia}}, \bibinfo {author} {\bibfnamefont
  {M.}~\bibnamefont {Auf~der Maur}}, \bibinfo {author} {\bibfnamefont
  {A.}~\bibnamefont {Liedl}}, \bibinfo {author} {\bibfnamefont
  {R.}~\bibnamefont {Larciprete}}, \bibinfo {author} {\bibfnamefont {D.~V.}\
  \bibnamefont {Kuznetsov}}, \bibinfo {author} {\bibfnamefont {D.}~\bibnamefont
  {Saranin}},\ and\ \bibinfo {author} {\bibfnamefont {A.}~\bibnamefont
  {Di~Carlo}},\ }\href {https://doi.org/10.1038/s41563-019-0478-1} {\bibfield
  {journal} {\bibinfo  {journal} {Nat. Mater.}\ }\textbf {\bibinfo {volume}
  {18}},\ \bibinfo {pages} {1228} (\bibinfo {year} {2019})}\BibitemShut
  {NoStop}%
\bibitem [{\citenamefont {Martynov}\ \emph {et~al.}(2021)\citenamefont
  {Martynov}, \citenamefont {Nazmitdinov}, \citenamefont {Gladyshev},\ and\
  \citenamefont {Moià-Pol}}]{Mar}%
  \BibitemOpen
  \bibfield  {author} {\bibinfo {author} {\bibfnamefont {Y.~B.}\ \bibnamefont
  {Martynov}}, \bibinfo {author} {\bibfnamefont {R.~G.}\ \bibnamefont
  {Nazmitdinov}}, \bibinfo {author} {\bibfnamefont {P.~P.}\ \bibnamefont
  {Gladyshev}},\ and\ \bibinfo {author} {\bibfnamefont {A.}~\bibnamefont
  {Moià-Pol}},\ }\href {https://doi.org/doi.org/10.1016/j.mencom.2021.07.007}
  {\bibfield  {journal} {\bibinfo  {journal} {Mendeleev Communications}\
  }\textbf {\bibinfo {volume} {31}},\ \bibinfo {pages} {459} (\bibinfo {year}
  {2021})}\BibitemShut {NoStop}%
\bibitem [{\citenamefont {Li}\ \emph {et~al.}(2021)\citenamefont {Li},
  \citenamefont {Wang}, \citenamefont {Yang}, \citenamefont {Pu}, \citenamefont
  {Gao}, \citenamefont {He}, \citenamefont {Cao}, \citenamefont {Han},\ and\
  \citenamefont {Li}}]{Li2021}%
  \BibitemOpen
  \bibfield  {author} {\bibinfo {author} {\bibfnamefont {T.}~\bibnamefont
  {Li}}, \bibinfo {author} {\bibfnamefont {S.}~\bibnamefont {Wang}}, \bibinfo
  {author} {\bibfnamefont {J.}~\bibnamefont {Yang}}, \bibinfo {author}
  {\bibfnamefont {X.}~\bibnamefont {Pu}}, \bibinfo {author} {\bibfnamefont
  {B.}~\bibnamefont {Gao}}, \bibinfo {author} {\bibfnamefont {Z.}~\bibnamefont
  {He}}, \bibinfo {author} {\bibfnamefont {Q.}~\bibnamefont {Cao}}, \bibinfo
  {author} {\bibfnamefont {J.}~\bibnamefont {Han}},\ and\ \bibinfo {author}
  {\bibfnamefont {X.}~\bibnamefont {Li}},\ }\bibfield  {journal} {\bibinfo
  {journal} {Nano Energy}\ }\textbf {\bibinfo {volume} {82}},\ \href
  {https://doi.org/10.1016/J.NANOEN.2021.105742} {10.1016/J.NANOEN.2021.105742}
  (\bibinfo {year} {2021})\BibitemShut {NoStop}%
\bibitem [{\citenamefont {Zhang}\ \emph {et~al.}(2021)\citenamefont {Zhang},
  \citenamefont {Wu}, \citenamefont {Wang}, \citenamefont {Pan}, \citenamefont
  {Zhang}, \citenamefont {Wang}, \citenamefont {Lan},\ and\ \citenamefont
  {Lin}}]{Zhang2021}%
  \BibitemOpen
  \bibfield  {author} {\bibinfo {author} {\bibfnamefont {X.}~\bibnamefont
  {Zhang}}, \bibinfo {author} {\bibfnamefont {J.}~\bibnamefont {Wu}}, \bibinfo
  {author} {\bibfnamefont {S.}~\bibnamefont {Wang}}, \bibinfo {author}
  {\bibfnamefont {W.}~\bibnamefont {Pan}}, \bibinfo {author} {\bibfnamefont
  {M.}~\bibnamefont {Zhang}}, \bibinfo {author} {\bibfnamefont
  {X.}~\bibnamefont {Wang}}, \bibinfo {author} {\bibfnamefont {Z.}~\bibnamefont
  {Lan}},\ and\ \bibinfo {author} {\bibfnamefont {J.}~\bibnamefont {Lin}},\
  }\bibfield  {journal} {\bibinfo  {journal} {ACS Applied Energy Materials}\
  }\textbf {\bibinfo {volume} {4}},\ \href
  {https://doi.org/10.1021/ACSAEM.1C01142} {10.1021/ACSAEM.1C01142} (\bibinfo
  {year} {2021})\BibitemShut {NoStop}%
\bibitem [{\citenamefont {Pereyra}\ \emph {et~al.}(2021)\citenamefont
  {Pereyra}, \citenamefont {Xie},\ and\ \citenamefont
  {Lira-Cantu}}]{Pereyra2021}%
  \BibitemOpen
  \bibfield  {author} {\bibinfo {author} {\bibfnamefont {C.}~\bibnamefont
  {Pereyra}}, \bibinfo {author} {\bibfnamefont {H.}~\bibnamefont {Xie}},\ and\
  \bibinfo {author} {\bibfnamefont {M.}~\bibnamefont {Lira-Cantu}},\ }\href
  {https://doi.org/10.1016/J.JECHEM.2021.01.037} {\bibfield  {journal}
  {\bibinfo  {journal} {Journal of Energy Chemistry}\ }\textbf {\bibinfo
  {volume} {60}},\ \bibinfo {pages} {599} (\bibinfo {year} {2021})}\BibitemShut
  {NoStop}%
\bibitem [{\citenamefont {Bonn{\'{i}}n-Ripoll}\ \emph
  {et~al.}(2019)\citenamefont {Bonn{\'{i}}n-Ripoll}, \citenamefont {Martynov},
  \citenamefont {Cardona}, \citenamefont {Nazmitdinov},\ and\ \citenamefont
  {Pujol-Nadal}}]{Bonnin-Ripoll2019}%
  \BibitemOpen
  \bibfield  {author} {\bibinfo {author} {\bibfnamefont {F.}~\bibnamefont
  {Bonn{\'{i}}n-Ripoll}}, \bibinfo {author} {\bibfnamefont {Y.~B.}\
  \bibnamefont {Martynov}}, \bibinfo {author} {\bibfnamefont {G.}~\bibnamefont
  {Cardona}}, \bibinfo {author} {\bibfnamefont {R.~G.}\ \bibnamefont
  {Nazmitdinov}},\ and\ \bibinfo {author} {\bibfnamefont {R.}~\bibnamefont
  {Pujol-Nadal}},\ }\href {https://doi.org/10.1016/j.solmat.2019.110050}
  {\bibfield  {journal} {\bibinfo  {journal} {Solar Energy Materials and Solar
  Cells}\ }\textbf {\bibinfo {volume} {200}},\ \bibinfo {pages} {110050}
  (\bibinfo {year} {2019})}\BibitemShut {NoStop}%
\bibitem [{\citenamefont {Deng}\ \emph {et~al.}(2019)\citenamefont {Deng},
  \citenamefont {Li}, \citenamefont {Deng},\ and\ \citenamefont
  {Li}}]{Deng2019}%
  \BibitemOpen
  \bibfield  {author} {\bibinfo {author} {\bibfnamefont {K.}~\bibnamefont
  {Deng}}, \bibinfo {author} {\bibfnamefont {L.}~\bibnamefont {Li}}, \bibinfo
  {author} {\bibfnamefont {K.}~\bibnamefont {Deng}},\ and\ \bibinfo {author}
  {\bibfnamefont {L.}~\bibnamefont {Li}},\ }\bibfield  {journal} {\bibinfo
  {journal} {Small Methods}\ }\href {https://doi.org/10.1002/smtd.201900150}
  {10.1002/smtd.201900150} (\bibinfo {year} {2019})\BibitemShut {NoStop}%
\bibitem [{\citenamefont {Haidari}(2019)}]{Haidari2019}%
  \BibitemOpen
  \bibfield  {author} {\bibinfo {author} {\bibfnamefont {G.}~\bibnamefont
  {Haidari}},\ }\bibfield  {journal} {\bibinfo  {journal} {AIP Advances}\
  }\href {https://doi.org/10.1063/1.5110495} {10.1063/1.5110495} (\bibinfo
  {year} {2019})\BibitemShut {NoStop}%
\bibitem [{\citenamefont {Mishra}\ and\ \citenamefont
  {Shukla}(2020)}]{Mishra2020}%
  \BibitemOpen
  \bibfield  {author} {\bibinfo {author} {\bibfnamefont {A.~K.}\ \bibnamefont
  {Mishra}}\ and\ \bibinfo {author} {\bibfnamefont {R.~K.}\ \bibnamefont
  {Shukla}},\ }\href {https://doi.org/10.1016/J.MATPR.2020.11.376} {\bibfield
  {journal} {\bibinfo  {journal} {Materials Today: Proceedings}\ }\textbf
  {\bibinfo {volume} {49}},\ \bibinfo {pages} {3181} (\bibinfo {year}
  {2020})}\BibitemShut {NoStop}%
\bibitem [{\citenamefont {Husainat}\ \emph {et~al.}(2020)\citenamefont
  {Husainat}, \citenamefont {Ali}, \citenamefont {Cofie}, \citenamefont
  {Attia}, \citenamefont {Fuller},\ and\ \citenamefont
  {Darwish}}]{Husainat2020}%
  \BibitemOpen
  \bibfield  {author} {\bibinfo {author} {\bibfnamefont {A.}~\bibnamefont
  {Husainat}}, \bibinfo {author} {\bibfnamefont {W.}~\bibnamefont {Ali}},
  \bibinfo {author} {\bibfnamefont {P.}~\bibnamefont {Cofie}}, \bibinfo
  {author} {\bibfnamefont {J.}~\bibnamefont {Attia}}, \bibinfo {author}
  {\bibfnamefont {J.}~\bibnamefont {Fuller}},\ and\ \bibinfo {author}
  {\bibfnamefont {A.}~\bibnamefont {Darwish}},\ }\href
  {https://doi.org/10.11648/J.AJOP.20200801.12} {\bibfield  {journal} {\bibinfo
   {journal} {American Journal of Optics and Photonics}\ }\textbf {\bibinfo
  {volume} {8}},\ \bibinfo {pages} {6} (\bibinfo {year} {2020})}\BibitemShut
  {NoStop}%
\bibitem [{\citenamefont {Al-Hattab}\ \emph {et~al.}(2021)\citenamefont
  {Al-Hattab}, \citenamefont {Moudou}, \citenamefont {Khenfouch}, \citenamefont
  {Bajjou}, \citenamefont {Chrafih},\ and\ \citenamefont
  {Rahmani}}]{Hattab2021}%
  \BibitemOpen
  \bibfield  {author} {\bibinfo {author} {\bibfnamefont {M.}~\bibnamefont
  {Al-Hattab}}, \bibinfo {author} {\bibfnamefont {L.}~\bibnamefont {Moudou}},
  \bibinfo {author} {\bibfnamefont {M.}~\bibnamefont {Khenfouch}}, \bibinfo
  {author} {\bibfnamefont {O.}~\bibnamefont {Bajjou}}, \bibinfo {author}
  {\bibfnamefont {Y.}~\bibnamefont {Chrafih}},\ and\ \bibinfo {author}
  {\bibfnamefont {K.}~\bibnamefont {Rahmani}},\ }\bibfield  {journal} {\bibinfo
   {journal} {Solar Energy}\ }\textbf {\bibinfo {volume} {227}},\ \href
  {https://doi.org/10.1016/J.SOLENER.2021.08.084}
  {10.1016/J.SOLENER.2021.08.084} (\bibinfo {year} {2021})\BibitemShut
  {NoStop}%
\bibitem [{\citenamefont {Hussain}\ \emph {et~al.}(2021)\citenamefont
  {Hussain}, \citenamefont {Riaz}, \citenamefont {Nowsherwan}, \citenamefont
  {Jahangir}, \citenamefont {Raza}, \citenamefont {Iqbal}, \citenamefont
  {Sadiq}, \citenamefont {Hussain},\ and\ \citenamefont
  {Naseem}}]{Hussain2021}%
  \BibitemOpen
  \bibfield  {author} {\bibinfo {author} {\bibfnamefont {S.~S.}\ \bibnamefont
  {Hussain}}, \bibinfo {author} {\bibfnamefont {S.}~\bibnamefont {Riaz}},
  \bibinfo {author} {\bibfnamefont {G.~A.}\ \bibnamefont {Nowsherwan}},
  \bibinfo {author} {\bibfnamefont {K.}~\bibnamefont {Jahangir}}, \bibinfo
  {author} {\bibfnamefont {A.}~\bibnamefont {Raza}}, \bibinfo {author}
  {\bibfnamefont {M.~J.}\ \bibnamefont {Iqbal}}, \bibinfo {author}
  {\bibfnamefont {I.}~\bibnamefont {Sadiq}}, \bibinfo {author} {\bibfnamefont
  {S.~M.}\ \bibnamefont {Hussain}},\ and\ \bibinfo {author} {\bibfnamefont
  {S.}~\bibnamefont {Naseem}},\ }\bibfield  {journal} {\bibinfo  {journal}
  {Journal of Renewable Energy}\ }\textbf {\bibinfo {volume} {2021}},\ \href
  {https://doi.org/10.1155/2021/6668687} {10.1155/2021/6668687} (\bibinfo
  {year} {2021})\BibitemShut {NoStop}%
\bibitem [{\citenamefont {Yadav}\ \emph {et~al.}(2022)\citenamefont {Yadav},
  \citenamefont {Pawar}, \citenamefont {Nandi}, \citenamefont {Neerugatti},
  \citenamefont {Kim}, \citenamefont {Cho},\ and\ \citenamefont
  {Heo}}]{Yadav2022}%
  \BibitemOpen
  \bibfield  {author} {\bibinfo {author} {\bibfnamefont {R.~K.}\ \bibnamefont
  {Yadav}}, \bibinfo {author} {\bibfnamefont {P.~S.}\ \bibnamefont {Pawar}},
  \bibinfo {author} {\bibfnamefont {R.}~\bibnamefont {Nandi}}, \bibinfo
  {author} {\bibfnamefont {K.~R.~E.}\ \bibnamefont {Neerugatti}}, \bibinfo
  {author} {\bibfnamefont {Y.~T.}\ \bibnamefont {Kim}}, \bibinfo {author}
  {\bibfnamefont {J.~Y.}\ \bibnamefont {Cho}},\ and\ \bibinfo {author}
  {\bibfnamefont {J.}~\bibnamefont {Heo}},\ }\bibfield  {journal} {\bibinfo
  {journal} {Solar Energy Materials and Solar Cells}\ }\textbf {\bibinfo
  {volume} {244}},\ \href {https://doi.org/10.1016/J.SOLMAT.2022.111835}
  {10.1016/J.SOLMAT.2022.111835} (\bibinfo {year} {2022})\BibitemShut {NoStop}%
\bibitem [{\citenamefont {Ahmad}\ \emph {et~al.}(2022)\citenamefont {Ahmad},
  \citenamefont {Khan}, \citenamefont {Khan},\ and\ \citenamefont
  {Kim}}]{Ahmad2022}%
  \BibitemOpen
  \bibfield  {author} {\bibinfo {author} {\bibfnamefont {K.}~\bibnamefont
  {Ahmad}}, \bibinfo {author} {\bibfnamefont {M.~Q.}\ \bibnamefont {Khan}},
  \bibinfo {author} {\bibfnamefont {R.~A.}\ \bibnamefont {Khan}},\ and\
  \bibinfo {author} {\bibfnamefont {H.}~\bibnamefont {Kim}},\ }\bibfield
  {journal} {\bibinfo  {journal} {Optical Materials}\ }\textbf {\bibinfo
  {volume} {128}},\ \href {https://doi.org/10.1016/J.OPTMAT.2022.112458}
  {10.1016/J.OPTMAT.2022.112458} (\bibinfo {year} {2022})\BibitemShut {NoStop}%
\bibitem [{\citenamefont {Karthick}\ \emph {et~al.}(2022)\citenamefont
  {Karthick}, \citenamefont {Velumani},\ and\ \citenamefont
  {Bouclé}}]{Karthick2022}%
  \BibitemOpen
  \bibfield  {author} {\bibinfo {author} {\bibfnamefont {S.}~\bibnamefont
  {Karthick}}, \bibinfo {author} {\bibfnamefont {S.}~\bibnamefont {Velumani}},\
  and\ \bibinfo {author} {\bibfnamefont {J.}~\bibnamefont {Bouclé}},\
  }\bibfield  {journal} {\bibinfo  {journal} {Optical Materials}\ }\textbf
  {\bibinfo {volume} {126}},\ \href
  {https://doi.org/10.1016/J.OPTMAT.2022.112250} {10.1016/J.OPTMAT.2022.112250}
  (\bibinfo {year} {2022})\BibitemShut {NoStop}%
\bibitem [{\citenamefont {Burgelman}\ \emph {et~al.}(2021)\citenamefont
  {Burgelman}, \citenamefont {Decock}, \citenamefont {Niemegeers},
  \citenamefont {Verschraegen},\ and\ \citenamefont {Degrave}}]{scaps1dmanual}%
  \BibitemOpen
  \bibfield  {author} {\bibinfo {author} {\bibfnamefont {M.}~\bibnamefont
  {Burgelman}}, \bibinfo {author} {\bibfnamefont {K.}~\bibnamefont {Decock}},
  \bibinfo {author} {\bibfnamefont {A.}~\bibnamefont {Niemegeers}}, \bibinfo
  {author} {\bibfnamefont {J.}~\bibnamefont {Verschraegen}},\ and\ \bibinfo
  {author} {\bibfnamefont {S.}~\bibnamefont {Degrave}},\ }\href@noop {} {\emph
  {\bibinfo {title} {SCAPS-1D Manual}}}\ (\bibinfo {year} {2021})\BibitemShut
  {NoStop}%
\bibitem [{\citenamefont {{Deepthi Jayan}}\ and\ \citenamefont
  {Sebastian}(2021)}]{DeepthiJayan2021}%
  \BibitemOpen
  \bibfield  {author} {\bibinfo {author} {\bibfnamefont {K.}~\bibnamefont
  {{Deepthi Jayan}}}\ and\ \bibinfo {author} {\bibfnamefont {V.}~\bibnamefont
  {Sebastian}},\ }\href {https://doi.org/10.1016/J.SOLENER.2021.01.058}
  {\bibfield  {journal} {\bibinfo  {journal} {Solar Energy}\ }\textbf {\bibinfo
  {volume} {217}},\ \bibinfo {pages} {40} (\bibinfo {year} {2021})}\BibitemShut
  {NoStop}%
\bibitem [{\citenamefont {Bonn{\'{i}}n-Ripoll}\ \emph
  {et~al.}(2021)\citenamefont {Bonn{\'{i}}n-Ripoll}, \citenamefont {Martynov},
  \citenamefont {Nazmitdinov}, \citenamefont {Cardona},\ and\ \citenamefont
  {Pujol-Nadal}}]{Bonnin-Ripoll2021}%
  \BibitemOpen
  \bibfield  {author} {\bibinfo {author} {\bibfnamefont {F.}~\bibnamefont
  {Bonn{\'{i}}n-Ripoll}}, \bibinfo {author} {\bibfnamefont {Y.~B.}\
  \bibnamefont {Martynov}}, \bibinfo {author} {\bibfnamefont {R.~G.}\
  \bibnamefont {Nazmitdinov}}, \bibinfo {author} {\bibfnamefont
  {G.}~\bibnamefont {Cardona}},\ and\ \bibinfo {author} {\bibfnamefont
  {R.}~\bibnamefont {Pujol-Nadal}},\ }\bibfield  {journal} {\bibinfo  {journal}
  {Physical Chemistry Chemical Physics}\ }\href
  {https://doi.org/10.1039/D1CP03313A} {10.1039/D1CP03313A} (\bibinfo {year}
  {2021})\BibitemShut {NoStop}%
\bibitem [{\citenamefont {Xie}\ \emph {et~al.}(2021)\citenamefont {Xie},
  \citenamefont {Wang}, \citenamefont {Chen}, \citenamefont {Pereyra},
  \citenamefont {Pols}, \citenamefont {Ga{\l}kowski}, \citenamefont {Anaya},
  \citenamefont {Fu}, \citenamefont {Jia}, \citenamefont {Tang}, \citenamefont
  {Kubicki}, \citenamefont {Agarwalla}, \citenamefont {Kim}, \citenamefont
  {Prochowicz}, \citenamefont {Borris{\'{e}}}, \citenamefont {Bonn},
  \citenamefont {Bao}, \citenamefont {Sun}, \citenamefont {Zakeeruddin},
  \citenamefont {Emsley}, \citenamefont {Arbiol}, \citenamefont {Gao},
  \citenamefont {Fu}, \citenamefont {Wang}, \citenamefont {Tielrooij},
  \citenamefont {Stranks}, \citenamefont {Tao}, \citenamefont {Gr{\"{a}}tzel},
  \citenamefont {Hagfeldt},\ and\ \citenamefont {Lira-Cantu}}]{Xie2021}%
  \BibitemOpen
  \bibfield  {author} {\bibinfo {author} {\bibfnamefont {H.}~\bibnamefont
  {Xie}}, \bibinfo {author} {\bibfnamefont {Z.}~\bibnamefont {Wang}}, \bibinfo
  {author} {\bibfnamefont {Z.}~\bibnamefont {Chen}}, \bibinfo {author}
  {\bibfnamefont {C.}~\bibnamefont {Pereyra}}, \bibinfo {author} {\bibfnamefont
  {M.}~\bibnamefont {Pols}}, \bibinfo {author} {\bibfnamefont {K.}~\bibnamefont
  {Ga{\l}kowski}}, \bibinfo {author} {\bibfnamefont {M.}~\bibnamefont {Anaya}},
  \bibinfo {author} {\bibfnamefont {S.}~\bibnamefont {Fu}}, \bibinfo {author}
  {\bibfnamefont {X.}~\bibnamefont {Jia}}, \bibinfo {author} {\bibfnamefont
  {P.}~\bibnamefont {Tang}}, \bibinfo {author} {\bibfnamefont {D.~J.}\
  \bibnamefont {Kubicki}}, \bibinfo {author} {\bibfnamefont {A.}~\bibnamefont
  {Agarwalla}}, \bibinfo {author} {\bibfnamefont {H.~S.}\ \bibnamefont {Kim}},
  \bibinfo {author} {\bibfnamefont {D.}~\bibnamefont {Prochowicz}}, \bibinfo
  {author} {\bibfnamefont {X.}~\bibnamefont {Borris{\'{e}}}}, \bibinfo {author}
  {\bibfnamefont {M.}~\bibnamefont {Bonn}}, \bibinfo {author} {\bibfnamefont
  {C.}~\bibnamefont {Bao}}, \bibinfo {author} {\bibfnamefont {X.}~\bibnamefont
  {Sun}}, \bibinfo {author} {\bibfnamefont {S.~M.}\ \bibnamefont
  {Zakeeruddin}}, \bibinfo {author} {\bibfnamefont {L.}~\bibnamefont {Emsley}},
  \bibinfo {author} {\bibfnamefont {J.}~\bibnamefont {Arbiol}}, \bibinfo
  {author} {\bibfnamefont {F.}~\bibnamefont {Gao}}, \bibinfo {author}
  {\bibfnamefont {F.}~\bibnamefont {Fu}}, \bibinfo {author} {\bibfnamefont
  {H.~I.}\ \bibnamefont {Wang}}, \bibinfo {author} {\bibfnamefont {K.~J.}\
  \bibnamefont {Tielrooij}}, \bibinfo {author} {\bibfnamefont {S.~D.}\
  \bibnamefont {Stranks}}, \bibinfo {author} {\bibfnamefont {S.}~\bibnamefont
  {Tao}}, \bibinfo {author} {\bibfnamefont {M.}~\bibnamefont {Gr{\"{a}}tzel}},
  \bibinfo {author} {\bibfnamefont {A.}~\bibnamefont {Hagfeldt}},\ and\
  \bibinfo {author} {\bibfnamefont {M.}~\bibnamefont {Lira-Cantu}},\ }\href
  {https://doi.org/10.1016/J.JOULE.2021.04.003} {\bibfield  {journal} {\bibinfo
   {journal} {Joule}\ }\textbf {\bibinfo {volume} {5}},\ \bibinfo {pages}
  {1246} (\bibinfo {year} {2021})}\BibitemShut {NoStop}%
\bibitem [{\citenamefont {Liu}\ \emph {et~al.}(2016)\citenamefont {Liu},
  \citenamefont {Huang}, \citenamefont {Wei}, \citenamefont {Zheng},
  \citenamefont {Xiao},\ and\ \citenamefont {Gong}}]{Liu2016}%
  \BibitemOpen
  \bibfield  {author} {\bibinfo {author} {\bibfnamefont {H.}~\bibnamefont
  {Liu}}, \bibinfo {author} {\bibfnamefont {Z.}~\bibnamefont {Huang}}, \bibinfo
  {author} {\bibfnamefont {S.}~\bibnamefont {Wei}}, \bibinfo {author}
  {\bibfnamefont {L.}~\bibnamefont {Zheng}}, \bibinfo {author} {\bibfnamefont
  {L.}~\bibnamefont {Xiao}},\ and\ \bibinfo {author} {\bibfnamefont
  {Q.}~\bibnamefont {Gong}},\ }\href {https://doi.org/10.1039/c5nr05207f}
  {\bibfield  {journal} {\bibinfo  {journal} {Nanoscale}\ }\textbf {\bibinfo
  {volume} {8}},\ \bibinfo {pages} {6209} (\bibinfo {year} {2016})}\BibitemShut
  {NoStop}%
\bibitem [{\citenamefont {Cardona}\ and\ \citenamefont
  {Pujol-Nadal}(2020)}]{Cardona2020}%
  \BibitemOpen
  \bibfield  {author} {\bibinfo {author} {\bibfnamefont {G.}~\bibnamefont
  {Cardona}}\ and\ \bibinfo {author} {\bibfnamefont {R.}~\bibnamefont
  {Pujol-Nadal}},\ }\href {https://doi.org/10.1371/journal.pone.0240735}
  {\bibfield  {journal} {\bibinfo  {journal} {PLOS ONE}\ }\textbf {\bibinfo
  {volume} {15}},\ \bibinfo {pages} {e0240735} (\bibinfo {year}
  {2020})}\BibitemShut {NoStop}%
\bibitem [{\citenamefont {Byrnes}(2017)}]{StevenByrnes}%
  \BibitemOpen
  \bibfield  {author} {\bibinfo {author} {\bibfnamefont {S.}~\bibnamefont
  {Byrnes}},\ }\href {https://pypi.python.org/pypi/tmm} {\bibinfo {title} {{tmm
  0.1.7 : Python Package Index}}} (\bibinfo {year} {2017})\BibitemShut
  {NoStop}%
\bibitem [{\citenamefont {Byrnes}(2020)}]{byrnes2020multilayer}%
  \BibitemOpen
  \bibfield  {author} {\bibinfo {author} {\bibfnamefont {S.~J.}\ \bibnamefont
  {Byrnes}},\ }\href@noop {} {\bibinfo {title} {Multilayer optical
  calculations}} (\bibinfo {year} {2020}),\ \Eprint
  {https://arxiv.org/abs/1603.02720} {arXiv:1603.02720 [physics.comp-ph]}
  \BibitemShut {NoStop}%
\bibitem [{Fre(2001)}]{FreeCAD}%
  \BibitemOpen
  \href {https://www.freecadweb.org/} {\bibinfo {title} {{FreeCAD: Your Own 3D
  Parametric Modeler}}} (\bibinfo {year} {2001})\BibitemShut {NoStop}%
\bibitem [{\citenamefont {Liu}\ \emph {et~al.}(2014)\citenamefont {Liu},
  \citenamefont {Zhu}, \citenamefont {Wei}, \citenamefont {Li}, \citenamefont
  {Lv}, \citenamefont {Yang}, \citenamefont {Zhang}, \citenamefont {Yao},\ and\
  \citenamefont {Dai}}]{Liu2014}%
  \BibitemOpen
  \bibfield  {author} {\bibinfo {author} {\bibfnamefont {F.}~\bibnamefont
  {Liu}}, \bibinfo {author} {\bibfnamefont {J.}~\bibnamefont {Zhu}}, \bibinfo
  {author} {\bibfnamefont {J.}~\bibnamefont {Wei}}, \bibinfo {author}
  {\bibfnamefont {Y.}~\bibnamefont {Li}}, \bibinfo {author} {\bibfnamefont
  {M.}~\bibnamefont {Lv}}, \bibinfo {author} {\bibfnamefont {S.}~\bibnamefont
  {Yang}}, \bibinfo {author} {\bibfnamefont {B.}~\bibnamefont {Zhang}},
  \bibinfo {author} {\bibfnamefont {J.}~\bibnamefont {Yao}},\ and\ \bibinfo
  {author} {\bibfnamefont {S.}~\bibnamefont {Dai}},\ }\href
  {https://doi.org/10.1063/1.4885367} {\bibfield  {journal} {\bibinfo
  {journal} {Appl. Phys. Lett.}\ }\textbf {\bibinfo {volume} {104}},\ \bibinfo
  {pages} {253508} (\bibinfo {year} {2014})}\BibitemShut {NoStop}%
\bibitem [{\citenamefont {Hima}(2019)}]{hima}%
  \BibitemOpen
  \bibfield  {author} {\bibinfo {author} {\bibfnamefont {A.}~\bibnamefont
  {Hima}},\ }\href {https://doi.org/10.47238/ijeca.v4i1.92} {\bibfield
  {journal} {\bibinfo  {journal} {Int. J. Energetica}\ }\textbf {\bibinfo
  {volume} {4}},\ \bibinfo {pages} {56} (\bibinfo {year} {2019})}\BibitemShut
  {NoStop}%
\bibitem [{\citenamefont {Zhang}\ \emph {et~al.}(2017)\citenamefont {Zhang},
  \citenamefont {Wang}, \citenamefont {Zhu}, \citenamefont {Pellet},
  \citenamefont {Luo}, \citenamefont {Yi}, \citenamefont {Liu}, \citenamefont
  {Liu}, \citenamefont {Wang}, \citenamefont {Li}, \citenamefont {Xiao},
  \citenamefont {Zakeeruddin}, \citenamefont {Bi},\ and\ \citenamefont
  {Grätzel}}]{fei}%
  \BibitemOpen
  \bibfield  {author} {\bibinfo {author} {\bibfnamefont {F.}~\bibnamefont
  {Zhang}}, \bibinfo {author} {\bibfnamefont {Z.}~\bibnamefont {Wang}},
  \bibinfo {author} {\bibfnamefont {H.}~\bibnamefont {Zhu}}, \bibinfo {author}
  {\bibfnamefont {N.}~\bibnamefont {Pellet}}, \bibinfo {author} {\bibfnamefont
  {J.}~\bibnamefont {Luo}}, \bibinfo {author} {\bibfnamefont {C.}~\bibnamefont
  {Yi}}, \bibinfo {author} {\bibfnamefont {X.}~\bibnamefont {Liu}}, \bibinfo
  {author} {\bibfnamefont {H.}~\bibnamefont {Liu}}, \bibinfo {author}
  {\bibfnamefont {S.}~\bibnamefont {Wang}}, \bibinfo {author} {\bibfnamefont
  {X.}~\bibnamefont {Li}}, \bibinfo {author} {\bibfnamefont {Y.}~\bibnamefont
  {Xiao}}, \bibinfo {author} {\bibfnamefont {S.~M.}\ \bibnamefont
  {Zakeeruddin}}, \bibinfo {author} {\bibfnamefont {D.}~\bibnamefont {Bi}},\
  and\ \bibinfo {author} {\bibfnamefont {M.}~\bibnamefont {Grätzel}},\ }\href
  {https://doi.org/10.1016/j.nanoen.2017.09.035} {\bibfield  {journal}
  {\bibinfo  {journal} {Nano Energy}\ }\textbf {\bibinfo {volume} {41}},\
  \bibinfo {pages} {469} (\bibinfo {year} {2017})}\BibitemShut {NoStop}%
\bibitem [{\citenamefont {Farhana}\ \emph {et~al.}(2017)\citenamefont
  {Farhana}, \citenamefont {Mahbub}, \citenamefont {Satter},\ and\
  \citenamefont {Ullah}}]{hind}%
  \BibitemOpen
  \bibfield  {author} {\bibinfo {author} {\bibfnamefont {A.}~\bibnamefont
  {Farhana}}, \bibinfo {author} {\bibfnamefont {R.}~\bibnamefont {Mahbub}},
  \bibinfo {author} {\bibfnamefont {S.~S.}\ \bibnamefont {Satter}},\ and\
  \bibinfo {author} {\bibfnamefont {S.~M.}\ \bibnamefont {Ullah}},\ }\href
  {https://doi.org/10.1155/2017/9846310} {\bibfield  {journal} {\bibinfo
  {journal} {Inter. J. Photoenergy}\ }\textbf {\bibinfo {volume} {2017}},\
  \bibinfo {pages} {9846310} (\bibinfo {year} {2017})}\BibitemShut {NoStop}%
\bibitem [{\citenamefont {Chen}\ \emph {et~al.}(2019)\citenamefont {Chen},
  \citenamefont {Gulo}, \citenamefont {Chao},\ and\ \citenamefont
  {Liu}}]{Chen2019}%
  \BibitemOpen
  \bibfield  {author} {\bibinfo {author} {\bibfnamefont {H.~W.}\ \bibnamefont
  {Chen}}, \bibinfo {author} {\bibfnamefont {D.~P.}\ \bibnamefont {Gulo}},
  \bibinfo {author} {\bibfnamefont {Y.~C.}\ \bibnamefont {Chao}},\ and\
  \bibinfo {author} {\bibfnamefont {H.~L.}\ \bibnamefont {Liu}},\ }\href
  {https://doi.org/10.1038/s41598-019-54636-7} {\bibfield  {journal} {\bibinfo
  {journal} {Scientific Reports 2019 9:1}\ }\textbf {\bibinfo {volume} {9}},\
  \bibinfo {pages} {1} (\bibinfo {year} {2019})}\BibitemShut {NoStop}%
\bibitem [{\citenamefont {Wang}\ \emph {et~al.}(2017)\citenamefont {Wang},
  \citenamefont {Deng}, \citenamefont {Wang}, \citenamefont {Dai},
  \citenamefont {Xing}, \citenamefont {Zhan}, \citenamefont {Lu}, \citenamefont
  {Xie}, \citenamefont {Huang},\ and\ \citenamefont {Zheng}}]{xin}%
  \BibitemOpen
  \bibfield  {author} {\bibinfo {author} {\bibfnamefont {X.}~\bibnamefont
  {Wang}}, \bibinfo {author} {\bibfnamefont {L.-L.}\ \bibnamefont {Deng}},
  \bibinfo {author} {\bibfnamefont {L.-Y.}\ \bibnamefont {Wang}}, \bibinfo
  {author} {\bibfnamefont {S.-M.}\ \bibnamefont {Dai}}, \bibinfo {author}
  {\bibfnamefont {Z.}~\bibnamefont {Xing}}, \bibinfo {author} {\bibfnamefont
  {X.-X.}\ \bibnamefont {Zhan}}, \bibinfo {author} {\bibfnamefont {X.-Z.}\
  \bibnamefont {Lu}}, \bibinfo {author} {\bibfnamefont {S.-Y.}\ \bibnamefont
  {Xie}}, \bibinfo {author} {\bibfnamefont {R.-B.}\ \bibnamefont {Huang}},\
  and\ \bibinfo {author} {\bibfnamefont {L.-S.}\ \bibnamefont {Zheng}},\ }\href
  {https://doi.org/10.1039/C6TA07541J} {\bibfield  {journal} {\bibinfo
  {journal} {J. Mater. Chem. A}\ }\textbf {\bibinfo {volume} {5}},\ \bibinfo
  {pages} {1706} (\bibinfo {year} {2017})}\BibitemShut {NoStop}%
\bibitem [{\citenamefont {Wojciechowski}\ \emph {et~al.}(2014)\citenamefont
  {Wojciechowski}, \citenamefont {Saliba}, \citenamefont {Leijtens},
  \citenamefont {Abate},\ and\ \citenamefont {Snaith}}]{woj}%
  \BibitemOpen
  \bibfield  {author} {\bibinfo {author} {\bibfnamefont {K.}~\bibnamefont
  {Wojciechowski}}, \bibinfo {author} {\bibfnamefont {M.}~\bibnamefont
  {Saliba}}, \bibinfo {author} {\bibfnamefont {T.}~\bibnamefont {Leijtens}},
  \bibinfo {author} {\bibfnamefont {A.}~\bibnamefont {Abate}},\ and\ \bibinfo
  {author} {\bibfnamefont {H.~J.}\ \bibnamefont {Snaith}},\ }\href
  {https://doi.org/10.1039/C3EE43707H} {\bibfield  {journal} {\bibinfo
  {journal} {Energy Environ. Sci.}\ }\textbf {\bibinfo {volume} {7}},\ \bibinfo
  {pages} {1142} (\bibinfo {year} {2014})}\BibitemShut {NoStop}%
\bibitem [{\citenamefont {Snaith}\ and\ \citenamefont {Grätzel}(2006)}]{sna}%
  \BibitemOpen
  \bibfield  {author} {\bibinfo {author} {\bibfnamefont {H.~J.}\ \bibnamefont
  {Snaith}}\ and\ \bibinfo {author} {\bibfnamefont {M.}~\bibnamefont
  {Grätzel}},\ }\href {https://doi.org/10.1063/1.2424552} {\bibfield
  {journal} {\bibinfo  {journal} {Appl. Phys. Lett.}\ }\textbf {\bibinfo
  {volume} {89}},\ \bibinfo {pages} {262114} (\bibinfo {year}
  {2006})}\BibitemShut {NoStop}%
\bibitem [{\citenamefont {Dutta}\ \emph {et~al.}(2015)\citenamefont {Dutta},
  \citenamefont {Leeladhar}, \citenamefont {Pandey}, \citenamefont {Thakur},\
  and\ \citenamefont {Pal}}]{dutta}%
  \BibitemOpen
  \bibfield  {author} {\bibinfo {author} {\bibfnamefont {S.}~\bibnamefont
  {Dutta}}, \bibinfo {author} {\bibnamefont {Leeladhar}}, \bibinfo {author}
  {\bibfnamefont {A.}~\bibnamefont {Pandey}}, \bibinfo {author} {\bibfnamefont
  {O.~P.}\ \bibnamefont {Thakur}},\ and\ \bibinfo {author} {\bibfnamefont
  {R.}~\bibnamefont {Pal}},\ }\href {https://doi.org/10.1116/1.4904978}
  {\bibfield  {journal} {\bibinfo  {journal} {J. Vac. Sci. and Tech. A}\
  }\textbf {\bibinfo {volume} {33}},\ \bibinfo {pages} {021507} (\bibinfo
  {year} {2015})}\BibitemShut {NoStop}%
\bibitem [{\citenamefont {Green}\ \emph {et~al.}(2014)\citenamefont {Green},
  \citenamefont {Ho-Baillie},\ and\ \citenamefont {Snaith}}]{martin}%
  \BibitemOpen
  \bibfield  {author} {\bibinfo {author} {\bibfnamefont {M.}~\bibnamefont
  {Green}}, \bibinfo {author} {\bibfnamefont {A.}~\bibnamefont {Ho-Baillie}},\
  and\ \bibinfo {author} {\bibfnamefont {H.}~\bibnamefont {Snaith}},\ }\href
  {https://doi.org/10.1038/nphoton.2014.134} {\bibfield  {journal} {\bibinfo
  {journal} {Nat. Photonics}\ }\textbf {\bibinfo {volume} {8}},\ \bibinfo
  {pages} {506} (\bibinfo {year} {2014})}\BibitemShut {NoStop}%
\bibitem [{\citenamefont {García-Cañadas}\ \emph {et~al.}(2006)\citenamefont
  {García-Cañadas}, \citenamefont {Fabregat-Santiago}, \citenamefont
  {Bolink}, \citenamefont {Palomares}, \citenamefont {Garcia-Belmonte},\ and\
  \citenamefont {Bisquert}}]{gar}%
  \BibitemOpen
  \bibfield  {author} {\bibinfo {author} {\bibfnamefont {J.}~\bibnamefont
  {García-Cañadas}}, \bibinfo {author} {\bibfnamefont {F.}~\bibnamefont
  {Fabregat-Santiago}}, \bibinfo {author} {\bibfnamefont {H.~J.}\ \bibnamefont
  {Bolink}}, \bibinfo {author} {\bibfnamefont {E.}~\bibnamefont {Palomares}},
  \bibinfo {author} {\bibfnamefont {G.}~\bibnamefont {Garcia-Belmonte}},\ and\
  \bibinfo {author} {\bibfnamefont {J.}~\bibnamefont {Bisquert}},\ }\href
  {https://doi.org/10.1016/j.synthmet.2006.06.006} {\bibfield  {journal}
  {\bibinfo  {journal} {Synthetic Metals}\ }\textbf {\bibinfo {volume} {156}},\
  \bibinfo {pages} {944} (\bibinfo {year} {2006})}\BibitemShut {NoStop}%
\bibitem [{\citenamefont {Ching-Prado}\ \emph {et~al.}(2018)\citenamefont
  {Ching-Prado}, \citenamefont {Watson},\ and\ \citenamefont
  {Miranda}}]{Ching-Prado2018}%
  \BibitemOpen
  \bibfield  {author} {\bibinfo {author} {\bibfnamefont {E.}~\bibnamefont
  {Ching-Prado}}, \bibinfo {author} {\bibfnamefont {A.}~\bibnamefont
  {Watson}},\ and\ \bibinfo {author} {\bibfnamefont {H.}~\bibnamefont
  {Miranda}},\ }\href {https://doi.org/10.1007/s10854-018-8795-8} {\bibfield
  {journal} {\bibinfo  {journal} {Journal of Materials Science: Materials in
  Electronics}\ }\textbf {\bibinfo {volume} {29}},\ \bibinfo {pages} {15299}
  (\bibinfo {year} {2018})}\BibitemShut {NoStop}%
\bibitem [{\citenamefont {Siefke}\ \emph {et~al.}(2016)\citenamefont {Siefke},
  \citenamefont {Kroker}, \citenamefont {Pfeiffer}, \citenamefont {Puffky},
  \citenamefont {Dietrich}, \citenamefont {Franta}, \citenamefont
  {Ohl{\'{i}}dal}, \citenamefont {Szeghalmi}, \citenamefont {Kley},\ and\
  \citenamefont {T{\"{u}}nnermann}}]{Siefke2016}%
  \BibitemOpen
  \bibfield  {author} {\bibinfo {author} {\bibfnamefont {T.}~\bibnamefont
  {Siefke}}, \bibinfo {author} {\bibfnamefont {S.}~\bibnamefont {Kroker}},
  \bibinfo {author} {\bibfnamefont {K.}~\bibnamefont {Pfeiffer}}, \bibinfo
  {author} {\bibfnamefont {O.}~\bibnamefont {Puffky}}, \bibinfo {author}
  {\bibfnamefont {K.}~\bibnamefont {Dietrich}}, \bibinfo {author}
  {\bibfnamefont {D.}~\bibnamefont {Franta}}, \bibinfo {author} {\bibfnamefont
  {I.}~\bibnamefont {Ohl{\'{i}}dal}}, \bibinfo {author} {\bibfnamefont
  {A.}~\bibnamefont {Szeghalmi}}, \bibinfo {author} {\bibfnamefont {E.-B.}\
  \bibnamefont {Kley}},\ and\ \bibinfo {author} {\bibfnamefont
  {A.}~\bibnamefont {T{\"{u}}nnermann}},\ }\href
  {https://doi.org/10.1002/adom.201600250} {\bibfield  {journal} {\bibinfo
  {journal} {Advanced Optical Materials}\ }\textbf {\bibinfo {volume} {4}},\
  \bibinfo {pages} {1780} (\bibinfo {year} {2016})}\BibitemShut {NoStop}%
\bibitem [{\citenamefont {Chen}\ \emph {et~al.}(2015)\citenamefont {Chen},
  \citenamefont {Hsiao}, \citenamefont {Chen}, \citenamefont {Kang},
  \citenamefont {Huang},\ and\ \citenamefont {Lin}}]{Chen2015}%
  \BibitemOpen
  \bibfield  {author} {\bibinfo {author} {\bibfnamefont {C.~W.}\ \bibnamefont
  {Chen}}, \bibinfo {author} {\bibfnamefont {S.~Y.}\ \bibnamefont {Hsiao}},
  \bibinfo {author} {\bibfnamefont {C.~Y.}\ \bibnamefont {Chen}}, \bibinfo
  {author} {\bibfnamefont {H.~W.}\ \bibnamefont {Kang}}, \bibinfo {author}
  {\bibfnamefont {Z.~Y.}\ \bibnamefont {Huang}},\ and\ \bibinfo {author}
  {\bibfnamefont {H.~W.}\ \bibnamefont {Lin}},\ }\href
  {https://doi.org/10.1039/C4TA05237D} {\bibfield  {journal} {\bibinfo
  {journal} {Journal of Materials Chemistry A}\ }\textbf {\bibinfo {volume}
  {3}},\ \bibinfo {pages} {9152} (\bibinfo {year} {2015})}\BibitemShut
  {NoStop}%
\bibitem [{\citenamefont {Nichelatti}(2002)}]{Nichelatti2002}%
  \BibitemOpen
  \bibfield  {author} {\bibinfo {author} {\bibfnamefont {E.}~\bibnamefont
  {Nichelatti}},\ }\href {https://doi.org/10.1088/1464-4258/4/4/306} {\bibfield
   {journal} {\bibinfo  {journal} {J. Opt. A: Pure Appl. Opt}\ }\textbf
  {\bibinfo {volume} {4}},\ \bibinfo {pages} {400} (\bibinfo {year}
  {2002})}\BibitemShut {NoStop}%
\bibitem [{\citenamefont {Kim}\ \emph {et~al.}(2021)\citenamefont {Kim},
  \citenamefont {Kim}, \citenamefont {Kwon}, \citenamefont {Kim}, \citenamefont
  {Do},\ and\ \citenamefont {Woo}}]{Kim2021}%
  \BibitemOpen
  \bibfield  {author} {\bibinfo {author} {\bibfnamefont {K.~P.}\ \bibnamefont
  {Kim}}, \bibinfo {author} {\bibfnamefont {W.~H.}\ \bibnamefont {Kim}},
  \bibinfo {author} {\bibfnamefont {S.~M.}\ \bibnamefont {Kwon}}, \bibinfo
  {author} {\bibfnamefont {J.~Y.}\ \bibnamefont {Kim}}, \bibinfo {author}
  {\bibfnamefont {Y.~S.}\ \bibnamefont {Do}},\ and\ \bibinfo {author}
  {\bibfnamefont {S.}~\bibnamefont {Woo}},\ }\bibfield  {journal} {\bibinfo
  {journal} {Nanomaterials 2021, Vol. 11, Page 1233}\ }\textbf {\bibinfo
  {volume} {11}},\ \href {https://doi.org/10.3390/NANO11051233}
  {10.3390/NANO11051233} (\bibinfo {year} {2021})\BibitemShut {NoStop}%
\bibitem [{\citenamefont {Widianto}\ \emph {et~al.}(2021)\citenamefont
  {Widianto}, \citenamefont {Shobih}, \citenamefont {Rosa}, \citenamefont
  {Triyana}, \citenamefont {Nursam},\ and\ \citenamefont
  {Santoso}}]{Widianto2021}%
  \BibitemOpen
  \bibfield  {author} {\bibinfo {author} {\bibfnamefont {E.}~\bibnamefont
  {Widianto}}, \bibinfo {author} {\bibnamefont {Shobih}}, \bibinfo {author}
  {\bibfnamefont {E.~S.}\ \bibnamefont {Rosa}}, \bibinfo {author}
  {\bibfnamefont {K.}~\bibnamefont {Triyana}}, \bibinfo {author} {\bibfnamefont
  {N.~M.}\ \bibnamefont {Nursam}},\ and\ \bibinfo {author} {\bibfnamefont
  {I.}~\bibnamefont {Santoso}},\ }\bibfield  {journal} {\bibinfo  {journal}
  {Optical Materials}\ }\textbf {\bibinfo {volume} {121}},\ \href
  {https://doi.org/10.1016/J.OPTMAT.2021.111584} {10.1016/J.OPTMAT.2021.111584}
  (\bibinfo {year} {2021})\BibitemShut {NoStop}%
\bibitem [{\citenamefont {Mehrabian}\ \emph {et~al.}(2023)\citenamefont
  {Mehrabian}, \citenamefont {Akhavan}, \citenamefont {Rabiee}, \citenamefont
  {Afshar},\ and\ \citenamefont {Zare}}]{Mehrabian2023}%
  \BibitemOpen
  \bibfield  {author} {\bibinfo {author} {\bibfnamefont {M.}~\bibnamefont
  {Mehrabian}}, \bibinfo {author} {\bibfnamefont {O.}~\bibnamefont {Akhavan}},
  \bibinfo {author} {\bibfnamefont {N.}~\bibnamefont {Rabiee}}, \bibinfo
  {author} {\bibfnamefont {E.~N.}\ \bibnamefont {Afshar}},\ and\ \bibinfo
  {author} {\bibfnamefont {E.~N.}\ \bibnamefont {Zare}},\ }\bibfield  {journal}
  {\bibinfo  {journal} {Environmental Science and Pollution Research}\ }\textbf
  {\bibinfo {volume} {30}},\ \href
  {https://doi.org/10.1007/S11356-023-26497-1/FIGURES/6}
  {10.1007/S11356-023-26497-1/FIGURES/6} (\bibinfo {year} {2023})\BibitemShut
  {NoStop}%
\bibitem [{\citenamefont {Hossain}\ \emph {et~al.}(2023)\citenamefont
  {Hossain}, \citenamefont {Toki}, \citenamefont {Kuddus}, \citenamefont
  {Rubel}, \citenamefont {Hossain}, \citenamefont {Bencherif}, \citenamefont
  {Rahman}, \citenamefont {Islam},\ and\ \citenamefont
  {Mushtaq}}]{Hossain2023}%
  \BibitemOpen
  \bibfield  {author} {\bibinfo {author} {\bibfnamefont {M.~K.}\ \bibnamefont
  {Hossain}}, \bibinfo {author} {\bibfnamefont {G.~F.}\ \bibnamefont {Toki}},
  \bibinfo {author} {\bibfnamefont {A.}~\bibnamefont {Kuddus}}, \bibinfo
  {author} {\bibfnamefont {M.~H.}\ \bibnamefont {Rubel}}, \bibinfo {author}
  {\bibfnamefont {M.~M.}\ \bibnamefont {Hossain}}, \bibinfo {author}
  {\bibfnamefont {H.}~\bibnamefont {Bencherif}}, \bibinfo {author}
  {\bibfnamefont {M.~F.}\ \bibnamefont {Rahman}}, \bibinfo {author}
  {\bibfnamefont {M.~R.}\ \bibnamefont {Islam}},\ and\ \bibinfo {author}
  {\bibfnamefont {M.}~\bibnamefont {Mushtaq}},\ }\bibfield  {journal} {\bibinfo
   {journal} {Scientific Reports 2023 13:1}\ }\textbf {\bibinfo {volume}
  {13}},\ \href {https://doi.org/10.1038/s41598-023-28506-2}
  {10.1038/s41598-023-28506-2} (\bibinfo {year} {2023})\BibitemShut {NoStop}%
\bibitem [{\citenamefont {Hutchinson}\ \emph {et~al.}(2010)\citenamefont
  {Hutchinson}, \citenamefont {Coquil}, \citenamefont {Navid},\ and\
  \citenamefont {Pilon}}]{Hutchinson2010}%
  \BibitemOpen
  \bibfield  {author} {\bibinfo {author} {\bibfnamefont {N.~J.}\ \bibnamefont
  {Hutchinson}}, \bibinfo {author} {\bibfnamefont {T.}~\bibnamefont {Coquil}},
  \bibinfo {author} {\bibfnamefont {A.}~\bibnamefont {Navid}},\ and\ \bibinfo
  {author} {\bibfnamefont {L.}~\bibnamefont {Pilon}},\ }\href
  {https://doi.org/10.1016/J.TSF.2009.08.048} {\bibfield  {journal} {\bibinfo
  {journal} {Thin Solid Films}\ }\textbf {\bibinfo {volume} {518}},\ \bibinfo
  {pages} {2141} (\bibinfo {year} {2010})}\BibitemShut {NoStop}%
\bibitem [{\citenamefont {Raoult}\ \emph {et~al.}(2019)\citenamefont {Raoult},
  \citenamefont {Bodeux}, \citenamefont {Jutteau}, \citenamefont {Rives},
  \citenamefont {Yaiche}, \citenamefont {Coutancier}, \citenamefont {Rousset},\
  and\ \citenamefont {Collin}}]{Raoult2019}%
  \BibitemOpen
  \bibfield  {author} {\bibinfo {author} {\bibfnamefont {E.}~\bibnamefont
  {Raoult}}, \bibinfo {author} {\bibfnamefont {R.}~\bibnamefont {Bodeux}},
  \bibinfo {author} {\bibfnamefont {S.}~\bibnamefont {Jutteau}}, \bibinfo
  {author} {\bibfnamefont {S.}~\bibnamefont {Rives}}, \bibinfo {author}
  {\bibfnamefont {A.}~\bibnamefont {Yaiche}}, \bibinfo {author} {\bibfnamefont
  {D.}~\bibnamefont {Coutancier}}, \bibinfo {author} {\bibfnamefont
  {J.}~\bibnamefont {Rousset}},\ and\ \bibinfo {author} {\bibfnamefont
  {S.}~\bibnamefont {Collin}},\ }in\ \href
  {https://doi.org/10.4229/EUPVSEC20192019-3BV.2.53} {\emph {\bibinfo
  {booktitle} {EU PVSEC Proceedings}}},\ Vol.\ \bibinfo {volume} {36th EU PVSEC
  2019}\ (\bibinfo {year} {2019})\ pp.\ \bibinfo {pages} {757--763}\BibitemShut
  {NoStop}%
\bibitem [{\citenamefont {Raoult}\ \emph {et~al.}(2022)\citenamefont {Raoult},
  \citenamefont {Bodeux}, \citenamefont {Jutteau}, \citenamefont {Rives},
  \citenamefont {Yaiche}, \citenamefont {Blaizot}, \citenamefont {Coutancier},
  \citenamefont {Rousset},\ and\ \citenamefont {Collin}}]{Raoult2022}%
  \BibitemOpen
  \bibfield  {author} {\bibinfo {author} {\bibfnamefont {E.}~\bibnamefont
  {Raoult}}, \bibinfo {author} {\bibfnamefont {R.}~\bibnamefont {Bodeux}},
  \bibinfo {author} {\bibfnamefont {S.}~\bibnamefont {Jutteau}}, \bibinfo
  {author} {\bibfnamefont {S.}~\bibnamefont {Rives}}, \bibinfo {author}
  {\bibfnamefont {A.}~\bibnamefont {Yaiche}}, \bibinfo {author} {\bibfnamefont
  {A.}~\bibnamefont {Blaizot}}, \bibinfo {author} {\bibfnamefont
  {D.}~\bibnamefont {Coutancier}}, \bibinfo {author} {\bibfnamefont
  {J.}~\bibnamefont {Rousset}},\ and\ \bibinfo {author} {\bibfnamefont
  {S.}~\bibnamefont {Collin}},\ }\href {https://doi.org/10.1364/OE.444698}
  {\bibfield  {journal} {\bibinfo  {journal} {Optics Express}\ }\textbf
  {\bibinfo {volume} {30}},\ \bibinfo {pages} {9604} (\bibinfo {year}
  {2022})}\BibitemShut {NoStop}%
\bibitem [{\citenamefont {Hern{\'{a}}ndez-Granados}\ \emph
  {et~al.}(2019)\citenamefont {Hern{\'{a}}ndez-Granados}, \citenamefont
  {Corpus-Mendoza}, \citenamefont {Moreno-Romero}, \citenamefont
  {Rodr{\'{i}}guez-Casta{\~{n}}eda}, \citenamefont {Pascoe-Sussoni},
  \citenamefont {Castelo-Gonz{\'{a}}lez}, \citenamefont {Menchaca-Campos},
  \citenamefont {Escorcia-Garc{\'{i}}a},\ and\ \citenamefont
  {Hu}}]{Hernandez-Granados2019}%
  \BibitemOpen
  \bibfield  {author} {\bibinfo {author} {\bibfnamefont {A.}~\bibnamefont
  {Hern{\'{a}}ndez-Granados}}, \bibinfo {author} {\bibfnamefont {A.~N.}\
  \bibnamefont {Corpus-Mendoza}}, \bibinfo {author} {\bibfnamefont {P.~M.}\
  \bibnamefont {Moreno-Romero}}, \bibinfo {author} {\bibfnamefont {C.~A.}\
  \bibnamefont {Rodr{\'{i}}guez-Casta{\~{n}}eda}}, \bibinfo {author}
  {\bibfnamefont {J.~E.}\ \bibnamefont {Pascoe-Sussoni}}, \bibinfo {author}
  {\bibfnamefont {O.~A.}\ \bibnamefont {Castelo-Gonz{\'{a}}lez}}, \bibinfo
  {author} {\bibfnamefont {E.~C.}\ \bibnamefont {Menchaca-Campos}}, \bibinfo
  {author} {\bibfnamefont {J.}~\bibnamefont {Escorcia-Garc{\'{i}}a}},\ and\
  \bibinfo {author} {\bibfnamefont {H.}~\bibnamefont {Hu}},\ }\href
  {https://doi.org/10.1016/j.optmat.2018.12.044} {\bibfield  {journal}
  {\bibinfo  {journal} {Optical Materials}\ }\textbf {\bibinfo {volume} {88}},\
  \bibinfo {pages} {695} (\bibinfo {year} {2019})}\BibitemShut {NoStop}%
\bibitem [{\citenamefont {Landau}\ \emph {et~al.}(1995)\citenamefont {Landau},
  \citenamefont {Lifshitz},\ and\ \citenamefont {Pitaevskii}}]{LL}%
  \BibitemOpen
  \bibfield  {author} {\bibinfo {author} {\bibfnamefont {L.}~\bibnamefont
  {Landau}}, \bibinfo {author} {\bibfnamefont {E.}~\bibnamefont {Lifshitz}},\
  and\ \bibinfo {author} {\bibfnamefont {L.}~\bibnamefont {Pitaevskii}},\
  }\href@noop {} {\emph {\bibinfo {title} {{Electrodynamics of Continuous
  Media: Volume 8}}}},\ Course of theoretical physics\ (\bibinfo  {publisher}
  {Elsevier Science},\ \bibinfo {year} {1995})\ p.\ \bibinfo {pages}
  {460}\BibitemShut {NoStop}%
\bibitem [{\citenamefont {Sellers}\ and\ \citenamefont
  {Seebauer}(2011)}]{Sellers2011}%
  \BibitemOpen
  \bibfield  {author} {\bibinfo {author} {\bibfnamefont {M.~C.}\ \bibnamefont
  {Sellers}}\ and\ \bibinfo {author} {\bibfnamefont {E.~G.}\ \bibnamefont
  {Seebauer}},\ }\href {https://doi.org/10.1016/J.TSF.2010.10.071} {\bibfield
  {journal} {\bibinfo  {journal} {Thin Solid Films}\ }\textbf {\bibinfo
  {volume} {519}},\ \bibinfo {pages} {2103} (\bibinfo {year}
  {2011})}\BibitemShut {NoStop}%
\bibitem [{\citenamefont {Herz}(2017)}]{Herz2017}%
  \BibitemOpen
  \bibfield  {author} {\bibinfo {author} {\bibfnamefont {L.~M.}\ \bibnamefont
  {Herz}},\ }\href {https://doi.org/10.1021/acsenergylett.7b00276} {\bibfield
  {journal} {\bibinfo  {journal} {ACS Energy Letters}\ }\textbf {\bibinfo
  {volume} {2}},\ \bibinfo {pages} {1539} (\bibinfo {year} {2017})}\BibitemShut
  {NoStop}%
\bibitem [{\citenamefont {Bos-Coenraad}\ \emph {et~al.}(2017)\citenamefont
  {Bos-Coenraad}, \citenamefont {Bunthof},\ and\ \citenamefont
  {Schermer}}]{Bos17}%
  \BibitemOpen
  \bibfield  {author} {\bibinfo {author} {\bibfnamefont {J.}~\bibnamefont
  {Bos-Coenraad}}, \bibinfo {author} {\bibfnamefont {L.}~\bibnamefont
  {Bunthof}},\ and\ \bibinfo {author} {\bibfnamefont {J.}~\bibnamefont
  {Schermer}},\ }\href {https://doi.org/10.1016/J.SOLENER.2017.07.003}
  {\bibfield  {journal} {\bibinfo  {journal} {Solar Energy}\ }\textbf {\bibinfo
  {volume} {155}},\ \bibinfo {pages} {1188} (\bibinfo {year}
  {2017})}\BibitemShut {NoStop}%
\bibitem [{\citenamefont {Cardona}\ and\ \citenamefont
  {Pujol-Nadal}(2018)}]{Cardona2018}%
  \BibitemOpen
  \bibfield  {author} {\bibinfo {author} {\bibfnamefont {G.}~\bibnamefont
  {Cardona}}\ and\ \bibinfo {author} {\bibfnamefont {R.}~\bibnamefont
  {Pujol-Nadal}},\ }\href {https://github.com/bielcardona/OTSun} {\bibinfo
  {title} {{OTSun Python Package}}} (\bibinfo {year} {2018})\BibitemShut
  {NoStop}%
\bibitem [{OTS(2018)}]{OTSunWebApp}%
  \BibitemOpen
  \href {http://otsun.uib.es/otsunwebapp} {\bibinfo {title} {{OTSunWebApp}}}
  (\bibinfo {year} {2018})\BibitemShut {NoStop}%
\bibitem [{\citenamefont {Bonnín-Ripoll}\ \emph {et~al.}(2018)\citenamefont
  {Bonnín-Ripoll}, \citenamefont {Cardona},\ and\ \citenamefont
  {Nadal-Pujol}}]{Tutorial2}%
  \BibitemOpen
  \bibfield  {author} {\bibinfo {author} {\bibfnamefont {F.}~\bibnamefont
  {Bonnín-Ripoll}}, \bibinfo {author} {\bibfnamefont {G.}~\bibnamefont
  {Cardona}},\ and\ \bibinfo {author} {\bibfnamefont {R.}~\bibnamefont
  {Nadal-Pujol}},\ }\href
  {https://github.com/bielcardona/OTSun/blob/master/OTSunWebApp/Tutorial_2_1e_OTSun_WebApp.pdf}
  {\bibinfo {title} {{Tutorial 2 OTSun Web App}}} (\bibinfo {year}
  {2018})\BibitemShut {NoStop}%
\bibitem [{\citenamefont {{M. Ball}}\ \emph {et~al.}(2015)\citenamefont
  {{M. Ball}}, \citenamefont {{D. Stranks}}, \citenamefont
  {{T. H{\"{o}}rantner}}, \citenamefont {{Sven H{\"{u}}ttner}}, \citenamefont
  {{Wei Zhang}}, \citenamefont {{W. Crossland}}, \citenamefont
  {{Ivan Ramirez}}, \citenamefont {{Moritz Riede}}, \citenamefont
  {{B. Johnston}}, \citenamefont {{H. Friend}},\ and\ \citenamefont
  {{J. Snaith}}}]{M.Ball2015}%
  \BibitemOpen
  \bibfield  {author} {\bibinfo {author} {\bibfnamefont {J.}~\bibnamefont
  {{M. Ball}}}, \bibinfo {author} {\bibfnamefont {S.}~\bibnamefont
  {{D. Stranks}}}, \bibinfo {author} {\bibfnamefont {M.}~\bibnamefont
  {{T. H{\"{o}}rantner}}}, \bibinfo {author} {\bibnamefont
  {{Sven H{\"{u}}ttner}}}, \bibinfo {author} {\bibnamefont {{Wei Zhang}}},
  \bibinfo {author} {\bibfnamefont {E.~J.}\ \bibnamefont {{W. Crossland}}},
  \bibinfo {author} {\bibnamefont {{Ivan Ramirez}}}, \bibinfo {author}
  {\bibnamefont {{Moritz Riede}}}, \bibinfo {author} {\bibfnamefont
  {M.}~\bibnamefont {{B. Johnston}}}, \bibinfo {author} {\bibfnamefont
  {R.}~\bibnamefont {{H. Friend}}},\ and\ \bibinfo {author} {\bibfnamefont
  {H.}~\bibnamefont {{J. Snaith}}},\ }\href
  {https://doi.org/10.1039/C4EE03224A} {\bibfield  {journal} {\bibinfo
  {journal} {Energy \& Environmental Science}\ }\textbf {\bibinfo {volume}
  {8}},\ \bibinfo {pages} {602} (\bibinfo {year} {2015})}\BibitemShut {NoStop}%
\end{thebibliography}%
\end{document}